\documentclass[aps,pra,graphicx,9pt,twocolumn,superscriptaddress]{revtex4-2}
\usepackage{amsmath,amscd,graphicx,graphicx,cases}
\usepackage{multirow,float,subfigure,appendix}
\usepackage{diagbox,makecell,blindtext,color,amssymb}
\usepackage[colorlinks,linkcolor=blue,urlcolor=blue,
anchorcolor=blue,citecolor=blue]{hyperref}
\usepackage{textcomp,booktabs,colortbl,easyReview}
\newcolumntype{C}{c<{\kern\tabcolsep}@{}}
\usepackage{times}
\usepackage{lineno}
\usepackage{xcolor,ulem,bm}
\usepackage{orcidlink}

\begin{document}
\title{Observation of photonic dynamics in dissipative quantum Rabi models}
\author{Wen Ning\orcidlink{0000-0001-7919-8811}}
\thanks{These authors contribute equally to this work.}
\author{Ri-Hua Zheng\orcidlink{0000-0002-1944-1573}}
\thanks{These authors contribute equally to this work.}
\email{E-mail: ruazheng@gmail.com}
\author{Jia-Hao L\"{u}\orcidlink{0009-0007-2578-2415}}
\author{Ken Chen\orcidlink{0009-0007-1849-5258}}
\author{Xin Zhu}
\author{Fan Wu\orcidlink{0000-0002-1279-2258}}
\author{Zhen-Biao Yang\orcidlink{0000-0002-3964-4714}}
\email{E-mail: zbyang@fzu.edu.cn}
\author{Shi-Biao Zheng\orcidlink{0000-0002-9405-4709}}
\email{E-mail: t96034@fzu.edu.cn}
\affiliation{Fujian Key Laboratory of Quantum Information and Quantum\\
Optics, College of Physics and Information Engineering, Fuzhou University, Fuzhou, Fujian 350108, China}

\begin{abstract}
The quantum Rabi model (QRM), composed of a qubit interacting with a
quantized photonic field, is a cornerstone of quantum optics. The QRM with
dominant unitary dynamics has been demonstrated in circuit quantum
electrodynamics (QED) systems, but an open QRM with a strong photonic
dissipation has not been experimentally explored. We here present the first
experimental demonstration of such an open system in circuit QED, featuring
a controlled competition between the coherent qubit-field interaction and
the photonic dissipation.
We map out the photon number distributions of the dissipative resonator for different coupling strengths in the steady state.
We further
observe the variation of the photon number during the system's evolution
towards the steady state with fixed control parameters. The results
demonstrate that the system's behaviors are significantly modified by the
photonic dissipation.
\end{abstract}

\maketitle

The Quantum Rabi model (QRM), which describes the interaction between a
qubit and a quantized field mode, lies at the heart of quantum optics \cite{Frisk2019,Forn2019}.
It reduces to the well-known Jaynes-Cummings model (JCM) \cite{Jaynes1963,Shore1993} in the
rotating-wave approximation, which is valid when the qubit-photon
interaction strength is much smaller than the frequency scale. When the
coupling is comparable to the frequencies, the counter-rotating wave terms
play a non-negligible role, resulting in a competition between the photonic
creation and annihilation associated with the excitation or deexcitation of
the qubit. This competition gives arise to the emergence of a cat-like
state, where two coherent states of the field with opposite phases are
entangled with the qubit's energy levels \cite{Zheng1998,Solano2003,Chen2021}. More intriguingly, it can exhibit
a superradiant phase transition \cite{Ashhab2013,Hwang2015,Shen2021,Cai2021,Zheng2022}, featuring a sharp increase of the
photon number near the critical point. In addition to fundamental interest,
the associated critical phenomena are useful for enhancement of the
sensitivity in quantum metrology \cite{Garbe2020,Chu2021,Lv2022,Zhu2023}.

Over the past few years, both the spectroscopic signatures \cite{Lv2018,Yoshihara2017} and
dynamical behaviors of the QRM have been experimentally explored in
different systems  \cite{Cai2021,Zheng2022,Yoshihara2018,Langford2017,Braumuller2017}. In most of these experiments, the
qubit-photon coupling strength is much stronger than the system dissipation,
so that the system evolution is dominated by the coherent dynamics. When the
photonic decaying rate is comparable with the ineraction strength, the
system may display new effects, e.g., the dissipative phase transition
\cite{Hwang2018,Zhu2020}. Recently, the dissipative QRM was demonstrated in an ion-trap
experiment \cite{Cai2022}, where the phononic mode of a trapped ion, which was coupled
to its electronic degree of freedom and subjected to an artificially
engineered reservoir, mimicked the photonic mode of the original QRM. However,
the open QRM with a naturally dissipative photonic mode has not been
reported so far.

We here present a demonstration of such a dissipative QRM in a circuit
quantum electrodynamics architecture, where a superconducting qubit is
coupled to the microwave field stored in a lossy microwave resonator by an
ac flux, which periodically modulates the qubit's frequency. This frequency
modulation mediates a sideband interaction between the qubit and the resonator,
with a controllable photonic swapping rate. A transverse drive transforms
this JCM into a QRM with a non-negligible photonic dissipation rate.
We investigate the photon-number distributions and average photon numbers for different effective coupling strengths after a long-time dynamics.
We further track the system evolutions for
fixed parameters. The results demonstrate that for each case the photon
number evolves towards a steady value without making oscillations, in
contrast with the unitary dynamics.

\begin{figure}
\centering
\includegraphics[width=8.6cm]{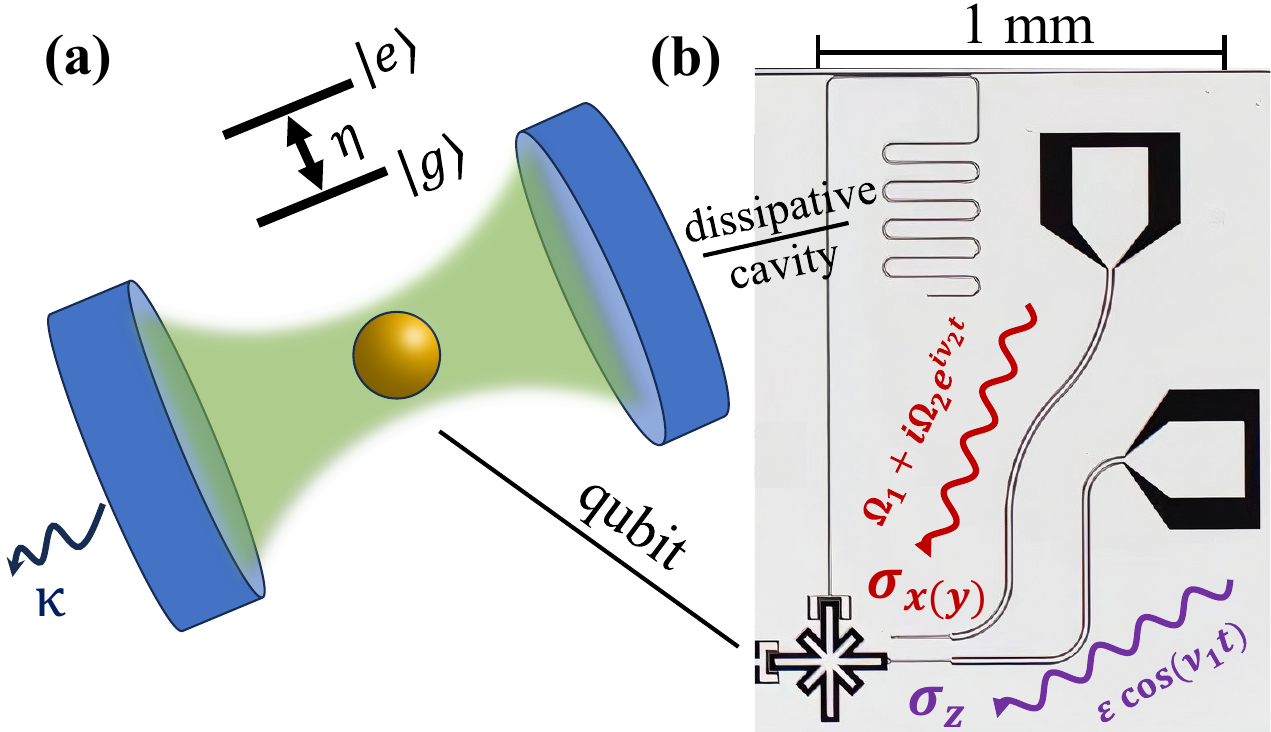}
\caption{\textbf{Experimental setup diagram.}
(a) Schematic diagram of the dissipative QRM, including a two-level artificial atom coupling to the field mode stored in a dissipative cavity (decay rate $\kappa \sim \eta$).
(b) On-chip demonstration of the dissipative QRM: optical micrograph of the superconducting circuit.
The Xmon qubit is capacitively coupled to a dissipative resonator and an XY control line [$\sigma_{x(y)}$], and inductively coupled to a Z control line ($\sigma_{z}$).
In addition to the dissipative resonator, the qubit is controllably coupled to a bus resonator with a negligible dissipation rate (not shown here), which is used to bring the qubit back to the ground state after the dissipative quantum Rabi dynamics, necessary for reading out the photon number of the dissipative resonator}
\label{fig1}
\end{figure}

The dissipative QRM [intuitively see Fig. \ref{fig1}(a)] dynamics can be described by the master equation ($\hbar=1$ hereafter and the decoherence of the qubit is ignored)
\begin{eqnarray}
\dot \rho=-i[H_{\rm Rabi},\rho]+\kappa a \rho a^\dag-\frac{\kappa}{2}(a^\dag a \rho +\rho a^\dag a),
\end{eqnarray}
where $\rho$ the density matrix, $a$ ($a^\dag$) the annihilation (creation) operator of the dissipative cavity mode with decay rate $\kappa$, and $H_{\rm Rabi}$ the QRM Hamiltonian
\begin{eqnarray} \label{QRM}
H_{\rm Rabi}= \frac{\Omega}{2} \sigma_y+ \delta a^\dag a +\eta \sigma_x (a+ a^\dag),
\end{eqnarray}
with effective frequencies $\Omega$ and $\delta$ of the qubit and cavity, respectively.
Note $\sigma_{x(y)}$ are the Pauli operators of the qubit under the ground and exited states basis \{$|g\rangle$, $|e\rangle$\} and $\sigma_y$ can be treated as $\sigma_z$ here, after a simple representation transformation (See supplemental material Sec. S1 for the details).
A distinct feature of the current model is that the cavity decay rate $\kappa=5$ MHz is close to the coupling strength $\eta\sim2\pi \times 1 $ MHz, results in a competition between the coherent QRM dynamics and incoherent photonic dissipation, and further leads to a steady state with a stable photon number (See supplemental material Sec. S2 for the detailed numerical simulation of system dynamics).

The experiment is carried out on an on-chip superconducting circuit \cite{song_2017_continuousvariable}, where the lowest two energy levels of an Xmon constitute the qubit, while a lossy microwave resonator acts as the dissipative cavity [see Fig. \ref{fig1}(b)].
Typically, this dissipative resonator is used to measure the qubit population through dispersive interaction \cite{Blais2004,Wallraff2004}, therefore with frequency $\omega_r/(2\pi)=6.656$ GHz much higher than the qubit frequency $\omega_q/(2\pi)=5.93\sim5.996$ GHz.
We utilize Floquet technology to couple the qubit and resonator at the first sideband.
Specifically, a periodical modulation, $\varepsilon \cos (\nu_1t)$, is applied through the Z line of the qubit [see Fig. \ref{fig1}(b)], resulting in a series of frequency splittings.
We adjust $\nu_1=\omega_r-\omega_q$ to resonate the qubit and resonator at the first sideband with strength $\eta=\lambda J_1(\varepsilon/\nu_1)/2$ [$J_1(\bullet)$ the first kind Bessel function at the first order], where $\lambda$ is the original interaction strength between the qubit and resonator.
Further, a transverse drive from the XY control line [see Fig. \ref{fig1}(b)], $\Omega_1+i\Omega_2 \exp(i \nu_2 t)$, transforms the JCM to the QRM in Eq. (\ref{QRM}) when $\nu_2=2\Omega_1 J_0(\varepsilon/\nu_1)$, with the effective frequency of resonator  $\Omega/2=\Omega_2J_0(\varepsilon/\nu_1)/2$ (see Supplemental Material Sec. S1 for the deviation).
By adjusting $\varepsilon$ and $\nu_1$, we can gradually change the value of $\eta/(2\pi)$ from 0 to 1 MHz with fixed frequencies of the qubit and resonator $\Omega/(2\pi)=1$ MHz and $\omega/(2\pi)=0.18$ MHz.
We first check the population oscillations of the dissipative QRM, realized with the choice $\varepsilon/(2\pi)=56.7$ MHz, $\nu_1/(2\pi)=708.7$ MHz, $\Omega_1/(2\pi)=20$ MHz, and $\Omega_2/(2\pi)=1$ MHz, corresponding to $\eta/(2\pi)=0.8$ MHz.

\begin{figure}
\centering
\includegraphics[width=8.6cm]{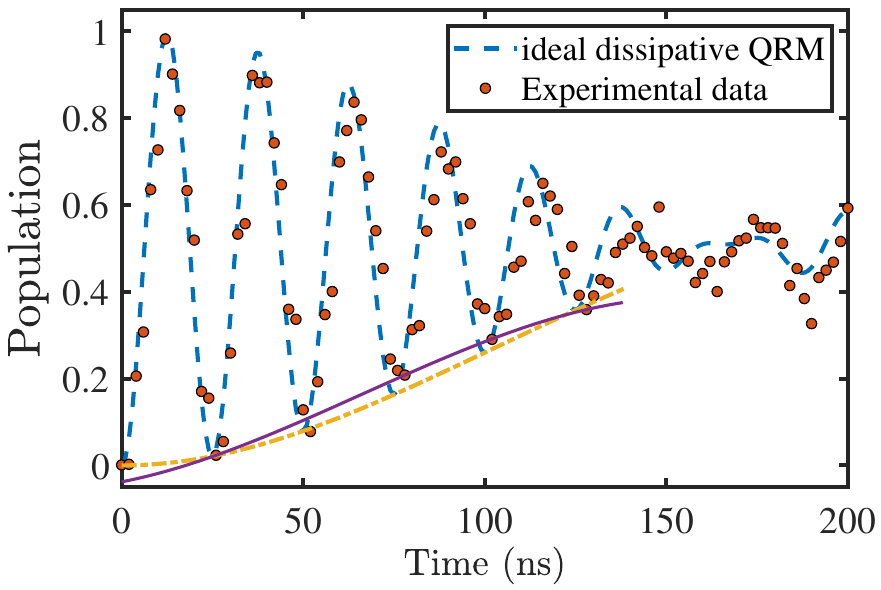}
\caption{\textbf{Observation of the $\bm{|e\rangle}$-state population evolution of the qubit. }
The effective dissipative Rabi model is realized for $\eta/(2\pi)=0.8$, $\omega/(2\pi)=0.18$, and $\Omega/(2\pi)=0.5$ MHz. The system starts with the ground state.
The line and dots denote the numerical result for the ideal dissipative QRM [considering the transverse field rotating $2\Omega_1 J_0(\varepsilon/\nu_1)\sigma_x$] and the experimental result, respectively.
The solid purple and black lines denote the lower and upper envelopes of the observed fast oscillations, respectively. These envelopes, respectively starting from the qubit's ground and excited states, are in well agreement with the qubit dynamics of the effective dissipative QRM with the corresponding initial states (dashed-dotted lines).}
\label{Rabicheck}
\end{figure}

The system starts from the ground state $|g\rangle_q\otimes|0\rangle_r$ (qubit's ground state and resonator's vacuum state).
After a preset interaction time, both the longitudinal modulation and transverse driving are switched off, so that the qubit is effectively decoupled from the dissipative resonator, as their detuning $(\omega_r-\omega_q)/(2\pi) \sim 700$ MHz is much larger than the coupling strength $\lambda/(2\pi)=40$ MHz. Then the qubit's state can be read out. Figure. \ref{Rabicheck} displays the measured population of the qubit's excited state.
We note that the fast oscillations, with the period of about $2\pi/[2\Omega_1 J_0(\varepsilon/\nu_1)]=25$ ns, are due to the transverse driving. In other words, the dissipative Rabi model is realized in the framework rotating at the frequency $2\Omega_1 J_0(\varepsilon/\nu_1)$ \cite{Zheng2022}, but the results are measured in the laboratory framework. Consequently, the envelopes of these oscillations reflect the qubit dynamics of the effective dissipative QRM \cite{Braumuller2017}. The upper and lower envelopes (solid lines) coincide the qubit excitation-numbers evolutions, starting from the initial excited and ground states, respectively. These agreements confirm the validity of the engineered dissipative QRM.
During the first several oscillatory periods, the experimental results are in well agreement with the simulation (line), confirming the validity of the approximations for deriving the effective dissipative QRM. With the elapse of time, the qubit becomes more and more affected by the dephasing noises, which destroy the osicllatory signals but are not included in the effective model.

\begin{figure}
\centering
\includegraphics[width=8cm]{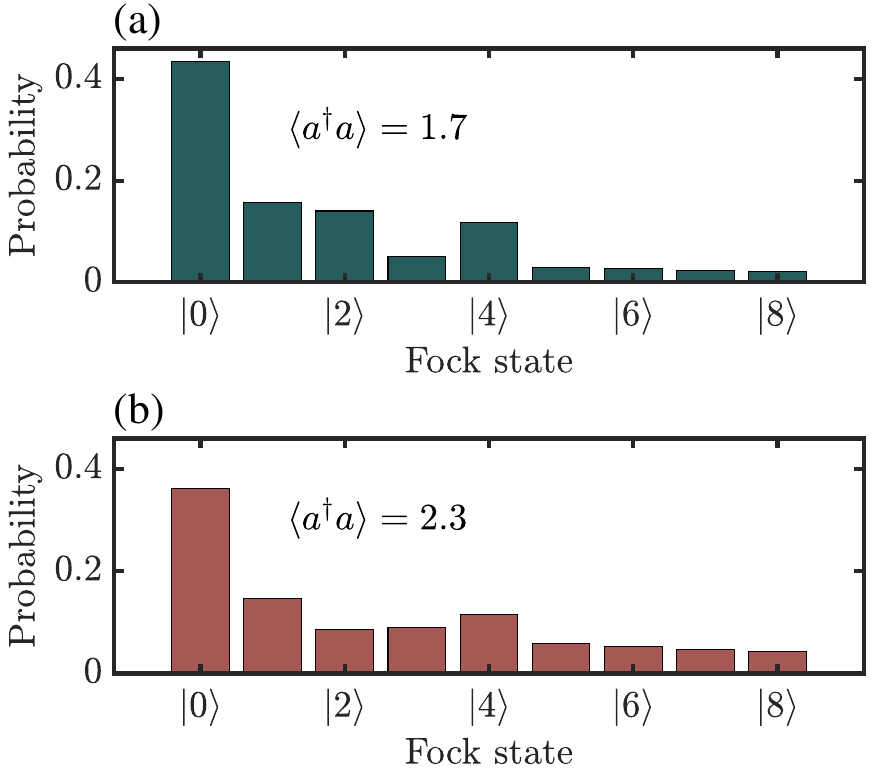}
\caption{\textbf{Observed photon number distributions for $\bm{\eta/(2\pi)=}$ 0.8 MHz (a) and 0.9 MHz (b)}. For each case, the photon number distribution is obtained after a dissipative QRM dynamics lasting for 3 $\mu$s. The result is measured with the qubit, whose excitation is transferred to the bus resonator after the quantum Rabi dynamics, following which it is coupled to the dissipative resonator to extract the photon-number populations.
%The smaller width bars represent the data for the numerical simulation of the effective dissipative QRM.
In (a), the measured $N$-photon populations for $N=0$ to 8 are 0.4348, 0.1566, 0.1401, 0.0511, 0.1170, 0.0292, 0.0263, 0.0236, and  0.0213, respectively.  In (b), the corresponding populations are 0.3629, 0.1460, 0.0866, 0.0898, 0.1156, 0.0579, 0.0521, 0.0469, and 0.0422, respectively.}
\label{photon_dis}
\end{figure}

The photon number of the dissipative resonator is measured with the help of the qubit. To do so, it is necessary to first transfer the excitation of the qubit to the bus resonator through a swapping gate at the frequency of the bus resonator. Following this excitation transfer, the qubit is biased back to the original frequency, where a longitudinal modulation is applied to mediate a resonant sideband interaction with the dissipative resonator, described by a dissipative JCM. The photon number distributions can be inferred from the Rabi signals of the qubit.
Figures \ref{photon_dis}(a) and \ref{photon_dis}(b) respectively display the photon number distributions for $\eta/(2\pi)=0.8$ and $0.9$ MHz, measured after a 3-$\mu$s dynamics of the dissipative QRM with the initial state $(|g\rangle_q+|e\rangle_q)\otimes |0\rangle_c/\sqrt{2}$. The results imply that the populations of relatively large photon numbers increases with the effective coupling strength.
%The hollow caps denote the results of the effective dissipative QRM.
For example, when $\eta/(2\pi)=0.8$ MHz the total population with three and more photons is 0.2684, which is increased to 0.4044 for $\eta/(2\pi)=0.9$ MHz.

%To characterize the photon-number distributions after a preset quench time, both the longitudinal modulation and transverse drive are switched off, effectively decoupling the qubit from the resonator since their detuning $(\omega_r-\omega_q)/(2\pi) \sim 700$ MHz is far larger than the coupling strength $\lambda/(2\pi)=40$ MHz.
%Subsequently, the qubit is tuned on resonance with the bus resonator [not shown in Fig. \ref{fig1}(b)] to remove the excitation caused by the quench dynamics. After this step, the qubit is in ground state $|g\rangle$ and can be further used to measure the Rabi oscillations signals through resonating the qubit again with the dissipative resonator at the first sideband. Therefore, the photon-number distribution of the resonator can be inferred from the measured Rabi oscillations signals \cite{Nature 459 546} (see Supplemental Material Sec. S4 for the details \cite{supp}). We place such characterization of the photon-number distributions in Fig. \ref{photon_dis}, with cases $\eta/(2\pi)=0.8$ MHz and $\eta/(2\pi)=0.9$ MHz after 3 ${\rm \mu}$s quench dynamics.

\begin{figure}
\centering
\includegraphics[width=8.6cm]{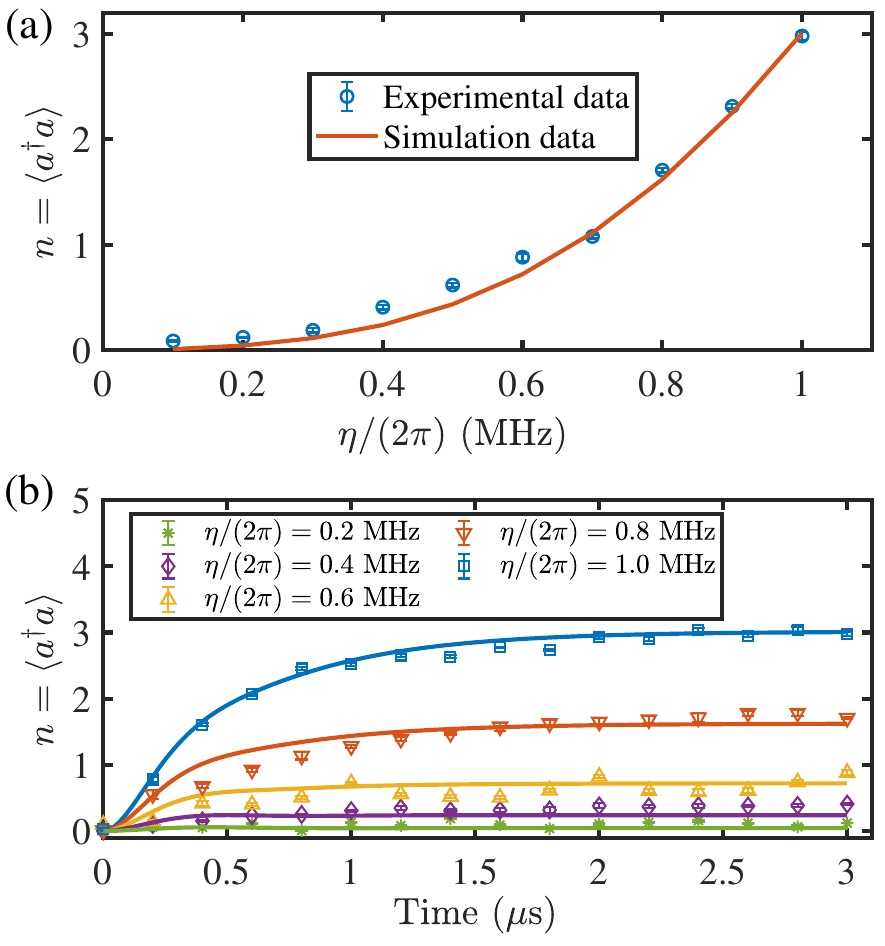}
\caption{\textbf{Observation of the dissipative QRM dynamics. }
(a) Photon number $n$ versus the QRM coupling $\eta/(2\pi)$.
Each data point is measured at $t=3$ $\rm \mu$s under the dissipative QRM dynamics with corresponding coupling $\eta/(2\pi)$.
(b) The variation of the photon number $n$ during the dissipative QRM system's evolution towards the steady state with fixed coupling $\eta/(2\pi)$.	}
\label{result}
\end{figure}

Pushing one step further, we investigate the average photon number in the steady state of the dissipative QRM versus the effective coupling strength $\eta$.
Figure. \ref{result}(a) display the results, all measured for the initial state $(|g\rangle_q+|e\rangle_q)\otimes |0\rangle_c/\sqrt{2}$, which evolves according to the dissipative QRM for an interaction time 3 $\mu$s. As expected, with the increase of $\eta$, the average photon number monotonously increases. These experimental results coincide the simulation (lines) very well. To confirm the system has well approached the steady state after 3 $\mu$s, we track the photon number evolutions within 3 $\mu$s. Figure. \ref{result}(b) shows the photon number evolutions observed for different values of $\eta$. As expected, for each case the photon number almost remains unchanged from 2 to 3 $\mu$s, where the average growth rates of the photons number are 0.0033, 0.0056, 0.0116, 0.0113, and 0.0088 ${\mu \rm{s}}^{-1}$ for $\eta/(2\pi)=0.2$, 0.4, 0.6, 0.8, and 1.0 MHz, respectively.
This proves that the cavity dissipation (with rate $\kappa=5$ MHz) plays an important role in the dynamics, making the system tend to  a steady state, in stark contrast with the oscillatory pattern exhibited by the unitary QRM under the time-independent Hamiltonian.
The results also demonstrate, the closer $\eta/(2\pi)$ to 1 MHz, the larger the changing rate of the steady-state photon number with respect to $\eta$. Due to the limitation of system parameters, the QRM is realized with an effective frequency ratio of 5.6. With the improvement of this ratio, it is possible to observe the dissipative superradiant phase transition, which was predicted to occur at the critical point $\xi_c=\sqrt{1+\kappa^2/(4\delta^2)}$ \cite{Hwang2018} when this ratio approaches infinity.

In conclusion, we have demonstrated the dynamics of the dissipative QRM engieered with a frequency-tunable Xmon qubit, together with its readout resonator. The qubit-resonator interaction is mediated at the first upper sideband with respect to a frequency modulation applied to the qubit, which makes the photonic swapping rate tunable. Thanks to this tunability, a transverse microwave drive transforms this JCM into a QRM, with a strong photonic dissipation, in a rotated framework. The effective qubit frequency in the QRM is controlled by the amplitude of a second transverse drive. The observed evolution of the qubit excited state population shows fast oscillations, with the envelopes well agreeing with the qubit dynamics in the effective dissipative QRM. The photon number distribution of the resonator after a preset interaction time is extracted by re-initiating the sideband interaction and recording the Rabi signal, governed by dissipative JCM. The observed average photon number monotonically increases with the effective coupling-frequency ratio, tending towards a steady value after a long-time dynamics, which makes the dissipative QRM different from the unitary counterpart. The method can be extended to synthesize a dissipative Dicke model involving multiple qubits coupled to a decaying resonator, with a controllable effective coupling-frequency ratio. We plan to investigate the dissipative superradiant phase transition in such a model.

The supplementary material includes the engineering of the Rabi model with controlled unitary-dissipative competition, simulation of the system dynamics, realization of the quench process, observation of the photon number evolution during the quench process, and steady-state photon-number distributions.

This work was supported by the National Natural Science Foundation of China under Grant Nos. 12274080, 12474356, and 12475015, and the Innovation Program for Quantum Science and Technology under Grant No. 2021ZD0300200.
\section*{AUTHOR DECLARATIONS}
\subsection*{Conflict of Interest}
The authors have no conflicts to disclose.
\subsection*{Author Contributions}

Wen Ning: Data curation (equal); Formal analysis (equal); Investigation (equal); Software (equal). Ri-Hua Zheng: Data curation (equal); Formal analysis (equal); Investigation (equal); Software (equal); Writing – review \& editing (equal). Jia-Hao L\"{u}: Data curation (equal); Visualization (equal). Ken Chen: Data curation (equal); Formal analysis (equal). Xin Zhu: Data curation (equal); Validation (equal). Fan Wu: Methodology (equal); Formal analysis (equal). Zhen-Biao Yang: Funding acquisition (equal); Project administration (equal); Supervision (equal); Writing – review \& editing (equal). Shi-Biao Zheng: Conceptualization (equal); Funding acquisition (equal); Supervision (equal); Writing – original draft (equal).
\section*{DATA AVAILABILITY}
The data that support the findings of this study are available within the article.
%#############################

\bibliography{ref}

\begin{thebibliography}{27}%
\makeatletter
\providecommand \@ifxundefined [1]{%
 \@ifx{#1\undefined}
}%
\providecommand \@ifnum [1]{%
 \ifnum #1\expandafter \@firstoftwo
 \else \expandafter \@secondoftwo
 \fi
}%
\providecommand \@ifx [1]{%
 \ifx #1\expandafter \@firstoftwo
 \else \expandafter \@secondoftwo
 \fi
}%
\providecommand \natexlab [1]{#1}%
\providecommand \enquote  [1]{``#1''}%
\providecommand \bibnamefont  [1]{#1}%
\providecommand \bibfnamefont [1]{#1}%
\providecommand \citenamefont [1]{#1}%
\providecommand \href@noop [0]{\@secondoftwo}%
\providecommand \href [0]{\begingroup \@sanitize@url \@href}%
\providecommand \@href[1]{\@@startlink{#1}\@@href}%
\providecommand \@@href[1]{\endgroup#1\@@endlink}%
\providecommand \@sanitize@url [0]{\catcode `\\12\catcode `\$12\catcode
  `\&12\catcode `\#12\catcode `\^12\catcode `\_12\catcode `\%12\relax}%
\providecommand \@@startlink[1]{}%
\providecommand \@@endlink[0]{}%
\providecommand \url  [0]{\begingroup\@sanitize@url \@url }%
\providecommand \@url [1]{\endgroup\@href {#1}{\urlprefix }}%
\providecommand \urlprefix  [0]{URL }%
\providecommand \Eprint [0]{\href }%
\providecommand \doibase [0]{https://doi.org/}%
\providecommand \selectlanguage [0]{\@gobble}%
\providecommand \bibinfo  [0]{\@secondoftwo}%
\providecommand \bibfield  [0]{\@secondoftwo}%
\providecommand \translation [1]{[#1]}%
\providecommand \BibitemOpen [0]{}%
\providecommand \bibitemStop [0]{}%
\providecommand \bibitemNoStop [0]{.\EOS\space}%
\providecommand \EOS [0]{\spacefactor3000\relax}%
\providecommand \BibitemShut  [1]{\csname bibitem#1\endcsname}%
\let\auto@bib@innerbib\@empty
%</preamble>
\bibitem [{\citenamefont {Frisk~Kockum}\ \emph {et~al.}(2019)\citenamefont
  {Frisk~Kockum}, \citenamefont {Miranowicz}, \citenamefont {De~Liberato},
  \citenamefont {Savasta},\ and\ \citenamefont {Nori}}]{Frisk2019}%
  \BibitemOpen
  \bibfield  {author} {\bibinfo {author} {\bibfnamefont {A.}~\bibnamefont
  {Frisk~Kockum}}, \bibinfo {author} {\bibfnamefont {A.}~\bibnamefont
  {Miranowicz}}, \bibinfo {author} {\bibfnamefont {S.}~\bibnamefont
  {De~Liberato}}, \bibinfo {author} {\bibfnamefont {S.}~\bibnamefont
  {Savasta}},\ and\ \bibinfo {author} {\bibfnamefont {F.}~\bibnamefont
  {Nori}},\ }\bibfield  {title} {\bibinfo {title} {Ultrastrong coupling between
  light and matter},\ }\href {https://doi.org/10.1038/s42254-018-0006-2}
  {\bibfield  {journal} {\bibinfo  {journal} {Nat. Rev. Phys.}\ }\textbf
  {\bibinfo {volume} {1}},\ \bibinfo {pages} {19} (\bibinfo {year}
  {2019})}\BibitemShut {NoStop}%
\bibitem [{\citenamefont {Forn-D\'{\i}az}\ \emph {et~al.}(2019)\citenamefont
  {Forn-D\'{\i}az}, \citenamefont {Lamata}, \citenamefont {Rico}, \citenamefont
  {Kono},\ and\ \citenamefont {Solano}}]{Forn2019}%
  \BibitemOpen
  \bibfield  {author} {\bibinfo {author} {\bibfnamefont {P.}~\bibnamefont
  {Forn-D\'{\i}az}}, \bibinfo {author} {\bibfnamefont {L.}~\bibnamefont
  {Lamata}}, \bibinfo {author} {\bibfnamefont {E.}~\bibnamefont {Rico}},
  \bibinfo {author} {\bibfnamefont {J.}~\bibnamefont {Kono}},\ and\ \bibinfo
  {author} {\bibfnamefont {E.}~\bibnamefont {Solano}},\ }\bibfield  {title}
  {\bibinfo {title} {Ultrastrong coupling regimes of light-matter
  interaction},\ }\href {https://doi.org/10.1103/RevModPhys.91.025005}
  {\bibfield  {journal} {\bibinfo  {journal} {Rev. Mod. Phys.}\ }\textbf
  {\bibinfo {volume} {91}},\ \bibinfo {pages} {025005} (\bibinfo {year}
  {2019})}\BibitemShut {NoStop}%
\bibitem [{\citenamefont {Jaynes}\ and\ \citenamefont
  {Cummings}(1963)}]{Jaynes1963}%
  \BibitemOpen
  \bibfield  {author} {\bibinfo {author} {\bibfnamefont {E.}~\bibnamefont
  {Jaynes}}\ and\ \bibinfo {author} {\bibfnamefont {F.}~\bibnamefont
  {Cummings}},\ }\bibfield  {title} {\bibinfo {title} {Comparison of quantum
  and semiclassical radiation theories with application to the beam maser},\
  }\href {https://doi.org/10.1109/PROC.1963.1664} {\bibfield  {journal}
  {\bibinfo  {journal} {Proc. IEEE}\ }\textbf {\bibinfo {volume} {51}},\
  \bibinfo {pages} {89} (\bibinfo {year} {1963})}\BibitemShut {NoStop}%
\bibitem [{\citenamefont {Shore}\ and\ \citenamefont
  {Knight}(1993)}]{Shore1993}%
  \BibitemOpen
  \bibfield  {author} {\bibinfo {author} {\bibfnamefont {B.~W.}\ \bibnamefont
  {Shore}}\ and\ \bibinfo {author} {\bibfnamefont {P.~L.}\ \bibnamefont
  {Knight}},\ }\bibfield  {title} {\bibinfo {title} {The {Jaynes}-{Cummings}
  {Model}},\ }\href {https://doi.org/10.1080/09500349314551321} {\bibfield
  {journal} {\bibinfo  {journal} {J. Mod. Opt.}\ }\textbf {\bibinfo {volume}
  {40}},\ \bibinfo {pages} {1195} (\bibinfo {year} {1993})}\BibitemShut
  {NoStop}%
\bibitem [{\citenamefont {Zheng}(1998)}]{Zheng1998}%
  \BibitemOpen
  \bibfield  {author} {\bibinfo {author} {\bibfnamefont {S.-B.}\ \bibnamefont
  {Zheng}},\ }\bibfield  {title} {\bibinfo {title} {Preparation of motional
  macroscopic quantum-interference states of a trapped ion},\ }\href
  {https://doi.org/10.1103/PhysRevA.58.761} {\bibfield  {journal} {\bibinfo
  {journal} {Phys. Rev. A}\ }\textbf {\bibinfo {volume} {58}},\ \bibinfo
  {pages} {761} (\bibinfo {year} {1998})}\BibitemShut {NoStop}%
\bibitem [{\citenamefont {Solano}\ \emph {et~al.}(2003)\citenamefont {Solano},
  \citenamefont {Agarwal},\ and\ \citenamefont {Walther}}]{Solano2003}%
  \BibitemOpen
  \bibfield  {author} {\bibinfo {author} {\bibfnamefont {E.}~\bibnamefont
  {Solano}}, \bibinfo {author} {\bibfnamefont {G.~S.}\ \bibnamefont
  {Agarwal}},\ and\ \bibinfo {author} {\bibfnamefont {H.}~\bibnamefont
  {Walther}},\ }\bibfield  {title} {\bibinfo {title} {Strong-driving-assisted
  multipartite entanglement in cavity {QED}},\ }\href
  {https://doi.org/10.1103/PhysRevLett.90.027903} {\bibfield  {journal}
  {\bibinfo  {journal} {Phys. Rev. Lett.}\ }\textbf {\bibinfo {volume} {90}},\
  \bibinfo {pages} {027903} (\bibinfo {year} {2003})}\BibitemShut {NoStop}%
\bibitem [{\citenamefont {Chen}\ \emph {et~al.}(2021)\citenamefont {Chen},
  \citenamefont {Qin}, \citenamefont {Wang}, \citenamefont {Miranowicz},\ and\
  \citenamefont {Nori}}]{Chen2021}%
  \BibitemOpen
  \bibfield  {author} {\bibinfo {author} {\bibfnamefont {Y.-H.}\ \bibnamefont
  {Chen}}, \bibinfo {author} {\bibfnamefont {W.}~\bibnamefont {Qin}}, \bibinfo
  {author} {\bibfnamefont {X.}~\bibnamefont {Wang}}, \bibinfo {author}
  {\bibfnamefont {A.}~\bibnamefont {Miranowicz}},\ and\ \bibinfo {author}
  {\bibfnamefont {F.}~\bibnamefont {Nori}},\ }\bibfield  {title} {\bibinfo
  {title} {Shortcuts to adiabaticity for the quantum {Rabi} model: Efficient
  generation of giant entangled cat states via parametric amplification},\
  }\href {https://doi.org/10.1103/PhysRevLett.126.023602} {\bibfield  {journal}
  {\bibinfo  {journal} {Phys. Rev. Lett.}\ }\textbf {\bibinfo {volume} {126}},\
  \bibinfo {pages} {023602} (\bibinfo {year} {2021})}\BibitemShut {NoStop}%
\bibitem [{\citenamefont {Ashhab}(2013)}]{Ashhab2013}%
  \BibitemOpen
  \bibfield  {author} {\bibinfo {author} {\bibfnamefont {S.}~\bibnamefont
  {Ashhab}},\ }\bibfield  {title} {\bibinfo {title} {Superradiance transition
  in a system with a single qubit and a single oscillator},\ }\href
  {https://doi.org/10.1103/PhysRevA.87.013826} {\bibfield  {journal} {\bibinfo
  {journal} {Phys. Rev. A}\ }\textbf {\bibinfo {volume} {87}},\ \bibinfo
  {pages} {013826} (\bibinfo {year} {2013})}\BibitemShut {NoStop}%
\bibitem [{\citenamefont {Hwang}\ \emph {et~al.}(2015)\citenamefont {Hwang},
  \citenamefont {Puebla},\ and\ \citenamefont {Plenio}}]{Hwang2015}%
  \BibitemOpen
  \bibfield  {author} {\bibinfo {author} {\bibfnamefont {M.-J.}\ \bibnamefont
  {Hwang}}, \bibinfo {author} {\bibfnamefont {R.}~\bibnamefont {Puebla}},\ and\
  \bibinfo {author} {\bibfnamefont {M.~B.}\ \bibnamefont {Plenio}},\ }\bibfield
   {title} {\bibinfo {title} {Quantum phase transition and universal dynamics
  in the {Rabi} model},\ }\href
  {https://doi.org/10.1103/PhysRevLett.115.180404} {\bibfield  {journal}
  {\bibinfo  {journal} {Phys. Rev. Lett.}\ }\textbf {\bibinfo {volume} {115}},\
  \bibinfo {pages} {180404} (\bibinfo {year} {2015})}\BibitemShut {NoStop}%
\bibitem [{\citenamefont {Shen}\ \emph {et~al.}(2021)\citenamefont {Shen},
  \citenamefont {Yang}, \citenamefont {Zhong}, \citenamefont {Yang},\ and\
  \citenamefont {Zheng}}]{Shen2021}%
  \BibitemOpen
  \bibfield  {author} {\bibinfo {author} {\bibfnamefont {L.-T.}\ \bibnamefont
  {Shen}}, \bibinfo {author} {\bibfnamefont {J.-W.}\ \bibnamefont {Yang}},
  \bibinfo {author} {\bibfnamefont {Z.-R.}\ \bibnamefont {Zhong}}, \bibinfo
  {author} {\bibfnamefont {Z.-B.}\ \bibnamefont {Yang}},\ and\ \bibinfo
  {author} {\bibfnamefont {S.-B.}\ \bibnamefont {Zheng}},\ }\bibfield  {title}
  {\bibinfo {title} {Quantum phase transition and quench dynamics in the
  two-mode {Rabi} model},\ }\href {https://doi.org/10.1103/PhysRevA.104.063703}
  {\bibfield  {journal} {\bibinfo  {journal} {Phys. Rev. A}\ }\textbf {\bibinfo
  {volume} {104}},\ \bibinfo {pages} {063703} (\bibinfo {year}
  {2021})}\BibitemShut {NoStop}%
\bibitem [{\citenamefont {Cai}\ \emph {et~al.}(2021)\citenamefont {Cai},
  \citenamefont {Liu}, \citenamefont {Zhao}, \citenamefont {Wu}, \citenamefont
  {Mei}, \citenamefont {Jiang}, \citenamefont {He}, \citenamefont {Zhang},
  \citenamefont {Zhou},\ and\ \citenamefont {Duan}}]{Cai2021}%
  \BibitemOpen
  \bibfield  {author} {\bibinfo {author} {\bibfnamefont {M.-L.}\ \bibnamefont
  {Cai}}, \bibinfo {author} {\bibfnamefont {Z.-D.}\ \bibnamefont {Liu}},
  \bibinfo {author} {\bibfnamefont {W.-D.}\ \bibnamefont {Zhao}}, \bibinfo
  {author} {\bibfnamefont {Y.-K.}\ \bibnamefont {Wu}}, \bibinfo {author}
  {\bibfnamefont {Q.-X.}\ \bibnamefont {Mei}}, \bibinfo {author} {\bibfnamefont
  {Y.}~\bibnamefont {Jiang}}, \bibinfo {author} {\bibfnamefont
  {L.}~\bibnamefont {He}}, \bibinfo {author} {\bibfnamefont {X.}~\bibnamefont
  {Zhang}}, \bibinfo {author} {\bibfnamefont {Z.-C.}\ \bibnamefont {Zhou}},\
  and\ \bibinfo {author} {\bibfnamefont {L.-M.}\ \bibnamefont {Duan}},\
  }\bibfield  {title} {\bibinfo {title} {Observation of a quantum phase
  transition in the quantum {Rabi} model with a single trapped ion},\ }\href
  {https://doi.org/10.1038/s41467-021-21425-8} {\bibfield  {journal} {\bibinfo
  {journal} {Nat. Commun.}\ }\textbf {\bibinfo {volume} {12}},\ \bibinfo
  {pages} {1126} (\bibinfo {year} {2021})}\BibitemShut {NoStop}%
\bibitem [{\citenamefont {Zheng}\ \emph {et~al.}(2023)\citenamefont {Zheng},
  \citenamefont {Ning}, \citenamefont {Chen}, \citenamefont {L\"u},
  \citenamefont {Shen}, \citenamefont {Xu}, \citenamefont {Zhang},
  \citenamefont {Xu}, \citenamefont {Li}, \citenamefont {Xia}, \citenamefont
  {Wu}, \citenamefont {Yang}, \citenamefont {Miranowicz}, \citenamefont
  {Lambert}, \citenamefont {Zheng}, \citenamefont {Fan}, \citenamefont {Nori},\
  and\ \citenamefont {Zheng}}]{Zheng2022}%
  \BibitemOpen
  \bibfield  {author} {\bibinfo {author} {\bibfnamefont {R.-H.}\ \bibnamefont
  {Zheng}}, \bibinfo {author} {\bibfnamefont {W.}~\bibnamefont {Ning}},
  \bibinfo {author} {\bibfnamefont {Y.-H.}\ \bibnamefont {Chen}}, \bibinfo
  {author} {\bibfnamefont {J.-H.}\ \bibnamefont {L\"u}}, \bibinfo {author}
  {\bibfnamefont {L.-T.}\ \bibnamefont {Shen}}, \bibinfo {author}
  {\bibfnamefont {K.}~\bibnamefont {Xu}}, \bibinfo {author} {\bibfnamefont
  {Y.-R.}\ \bibnamefont {Zhang}}, \bibinfo {author} {\bibfnamefont
  {D.}~\bibnamefont {Xu}}, \bibinfo {author} {\bibfnamefont {H.}~\bibnamefont
  {Li}}, \bibinfo {author} {\bibfnamefont {Y.}~\bibnamefont {Xia}}, \bibinfo
  {author} {\bibfnamefont {F.}~\bibnamefont {Wu}}, \bibinfo {author}
  {\bibfnamefont {Z.-B.}\ \bibnamefont {Yang}}, \bibinfo {author}
  {\bibfnamefont {A.}~\bibnamefont {Miranowicz}}, \bibinfo {author}
  {\bibfnamefont {N.}~\bibnamefont {Lambert}}, \bibinfo {author} {\bibfnamefont
  {D.}~\bibnamefont {Zheng}}, \bibinfo {author} {\bibfnamefont
  {H.}~\bibnamefont {Fan}}, \bibinfo {author} {\bibfnamefont {F.}~\bibnamefont
  {Nori}},\ and\ \bibinfo {author} {\bibfnamefont {S.-B.}\ \bibnamefont
  {Zheng}},\ }\bibfield  {title} {\bibinfo {title} {Observation of a
  superradiant phase transition with emergent cat states},\ }\href
  {https://doi.org/10.1103/PhysRevLett.131.113601} {\bibfield  {journal}
  {\bibinfo  {journal} {Phys. Rev. Lett.}\ }\textbf {\bibinfo {volume} {131}},\
  \bibinfo {pages} {113601} (\bibinfo {year} {2023})}\BibitemShut {NoStop}%
\bibitem [{\citenamefont {Garbe}\ \emph {et~al.}(2020)\citenamefont {Garbe},
  \citenamefont {Bina}, \citenamefont {Keller}, \citenamefont {Paris},\ and\
  \citenamefont {Felicetti}}]{Garbe2020}%
  \BibitemOpen
  \bibfield  {author} {\bibinfo {author} {\bibfnamefont {L.}~\bibnamefont
  {Garbe}}, \bibinfo {author} {\bibfnamefont {M.}~\bibnamefont {Bina}},
  \bibinfo {author} {\bibfnamefont {A.}~\bibnamefont {Keller}}, \bibinfo
  {author} {\bibfnamefont {M.~G.~A.}\ \bibnamefont {Paris}},\ and\ \bibinfo
  {author} {\bibfnamefont {S.}~\bibnamefont {Felicetti}},\ }\bibfield  {title}
  {\bibinfo {title} {Critical quantum metrology with a finite-component quantum
  phase transition},\ }\href {https://doi.org/10.1103/PhysRevLett.124.120504}
  {\bibfield  {journal} {\bibinfo  {journal} {Phys. Rev. Lett.}\ }\textbf
  {\bibinfo {volume} {124}},\ \bibinfo {pages} {120504} (\bibinfo {year}
  {2020})}\BibitemShut {NoStop}%
\bibitem [{\citenamefont {Chu}\ \emph {et~al.}(2021)\citenamefont {Chu},
  \citenamefont {Zhang}, \citenamefont {Yu},\ and\ \citenamefont
  {Cai}}]{Chu2021}%
  \BibitemOpen
  \bibfield  {author} {\bibinfo {author} {\bibfnamefont {Y.}~\bibnamefont
  {Chu}}, \bibinfo {author} {\bibfnamefont {S.}~\bibnamefont {Zhang}}, \bibinfo
  {author} {\bibfnamefont {B.}~\bibnamefont {Yu}},\ and\ \bibinfo {author}
  {\bibfnamefont {J.}~\bibnamefont {Cai}},\ }\bibfield  {title} {\bibinfo
  {title} {Dynamic framework for criticality-enhanced quantum sensing},\ }\href
  {https://doi.org/10.1103/PhysRevLett.126.010502} {\bibfield  {journal}
  {\bibinfo  {journal} {Phys. Rev. Lett.}\ }\textbf {\bibinfo {volume} {126}},\
  \bibinfo {pages} {010502} (\bibinfo {year} {2021})}\BibitemShut {NoStop}%
\bibitem [{\citenamefont {L\"u}\ \emph {et~al.}(2022)\citenamefont {L\"u},
  \citenamefont {Ning}, \citenamefont {Zhu}, \citenamefont {Wu}, \citenamefont
  {Shen}, \citenamefont {Yang},\ and\ \citenamefont {Zheng}}]{Lv2022}%
  \BibitemOpen
  \bibfield  {author} {\bibinfo {author} {\bibfnamefont {J.-H.}\ \bibnamefont
  {L\"u}}, \bibinfo {author} {\bibfnamefont {W.}~\bibnamefont {Ning}}, \bibinfo
  {author} {\bibfnamefont {X.}~\bibnamefont {Zhu}}, \bibinfo {author}
  {\bibfnamefont {F.}~\bibnamefont {Wu}}, \bibinfo {author} {\bibfnamefont
  {L.-T.}\ \bibnamefont {Shen}}, \bibinfo {author} {\bibfnamefont {Z.-B.}\
  \bibnamefont {Yang}},\ and\ \bibinfo {author} {\bibfnamefont {S.-B.}\
  \bibnamefont {Zheng}},\ }\bibfield  {title} {\bibinfo {title} {Critical
  quantum sensing based on the {Jaynes-Cummings} model with a squeezing
  drive},\ }\href {https://doi.org/10.1103/PhysRevA.106.062616} {\bibfield
  {journal} {\bibinfo  {journal} {Phys. Rev. A}\ }\textbf {\bibinfo {volume}
  {106}},\ \bibinfo {pages} {062616} (\bibinfo {year} {2022})}\BibitemShut
  {NoStop}%
\bibitem [{\citenamefont {Zhu}\ \emph {et~al.}(2023)\citenamefont {Zhu},
  \citenamefont {L\"u}, \citenamefont {Ning}, \citenamefont {Wu}, \citenamefont
  {Shen}, \citenamefont {Yang},\ and\ \citenamefont {Zheng}}]{Zhu2023}%
  \BibitemOpen
  \bibfield  {author} {\bibinfo {author} {\bibfnamefont {X.}~\bibnamefont
  {Zhu}}, \bibinfo {author} {\bibfnamefont {J.-H.}\ \bibnamefont {L\"u}},
  \bibinfo {author} {\bibfnamefont {W.}~\bibnamefont {Ning}}, \bibinfo {author}
  {\bibfnamefont {F.}~\bibnamefont {Wu}}, \bibinfo {author} {\bibfnamefont
  {L.-T.}\ \bibnamefont {Shen}}, \bibinfo {author} {\bibfnamefont {Z.-B.}\
  \bibnamefont {Yang}},\ and\ \bibinfo {author} {\bibfnamefont {S.-B.}\
  \bibnamefont {Zheng}},\ }\bibfield  {title} {\bibinfo {title}
  {Criticality-enhanced quantum sensing in the anisotropic quantum {Rabi}
  model},\ }\href {https://doi.org/10.1007/s11433-022-2073-9} {\bibfield
  {journal} {\bibinfo  {journal} {Sci. China-Phys. Mech. Astron.}\ }\textbf
  {\bibinfo {volume} {66}},\ \bibinfo {pages} {250313} (\bibinfo {year}
  {2023})}\BibitemShut {NoStop}%
\bibitem [{\citenamefont {Lv}\ \emph {et~al.}(2018)\citenamefont {Lv},
  \citenamefont {An}, \citenamefont {Liu}, \citenamefont {Zhang}, \citenamefont
  {Pedernales}, \citenamefont {Lamata}, \citenamefont {Solano},\ and\
  \citenamefont {Kim}}]{Lv2018}%
  \BibitemOpen
  \bibfield  {author} {\bibinfo {author} {\bibfnamefont {D.}~\bibnamefont
  {Lv}}, \bibinfo {author} {\bibfnamefont {S.}~\bibnamefont {An}}, \bibinfo
  {author} {\bibfnamefont {Z.}~\bibnamefont {Liu}}, \bibinfo {author}
  {\bibfnamefont {J.-N.}\ \bibnamefont {Zhang}}, \bibinfo {author}
  {\bibfnamefont {J.~S.}\ \bibnamefont {Pedernales}}, \bibinfo {author}
  {\bibfnamefont {L.}~\bibnamefont {Lamata}}, \bibinfo {author} {\bibfnamefont
  {E.}~\bibnamefont {Solano}},\ and\ \bibinfo {author} {\bibfnamefont
  {K.}~\bibnamefont {Kim}},\ }\bibfield  {title} {\bibinfo {title} {Quantum
  simulation of the quantum {Rabi Model} in a trapped ion},\ }\href
  {https://doi.org/10.1103/PhysRevX.8.021027} {\bibfield  {journal} {\bibinfo
  {journal} {Phys. Rev. X}\ }\textbf {\bibinfo {volume} {8}},\ \bibinfo {pages}
  {021027} (\bibinfo {year} {2018})}\BibitemShut {NoStop}%
\bibitem [{\citenamefont {Yoshihara}\ \emph {et~al.}(2017)\citenamefont
  {Yoshihara}, \citenamefont {Fuse}, \citenamefont {Ashhab}, \citenamefont
  {Kakuyanagi}, \citenamefont {Saito},\ and\ \citenamefont
  {Semba}}]{Yoshihara2017}%
  \BibitemOpen
  \bibfield  {author} {\bibinfo {author} {\bibfnamefont {F.}~\bibnamefont
  {Yoshihara}}, \bibinfo {author} {\bibfnamefont {T.}~\bibnamefont {Fuse}},
  \bibinfo {author} {\bibfnamefont {S.}~\bibnamefont {Ashhab}}, \bibinfo
  {author} {\bibfnamefont {K.}~\bibnamefont {Kakuyanagi}}, \bibinfo {author}
  {\bibfnamefont {S.}~\bibnamefont {Saito}},\ and\ \bibinfo {author}
  {\bibfnamefont {K.}~\bibnamefont {Semba}},\ }\bibfield  {title} {\bibinfo
  {title} {Superconducting qubit–oscillator circuit beyond the
  ultrastrong-coupling regime},\ }\href {https://doi.org/10.1038/nphys3906}
  {\bibfield  {journal} {\bibinfo  {journal} {Nat. Phys.}\ }\textbf {\bibinfo
  {volume} {13}},\ \bibinfo {pages} {44} (\bibinfo {year} {2017})}\BibitemShut
  {NoStop}%
\bibitem [{\citenamefont {Yoshihara}\ \emph {et~al.}(2018)\citenamefont
  {Yoshihara}, \citenamefont {Fuse}, \citenamefont {Ao}, \citenamefont
  {Ashhab}, \citenamefont {Kakuyanagi}, \citenamefont {Saito}, \citenamefont
  {Aoki}, \citenamefont {Koshino},\ and\ \citenamefont
  {Semba}}]{Yoshihara2018}%
  \BibitemOpen
  \bibfield  {author} {\bibinfo {author} {\bibfnamefont {F.}~\bibnamefont
  {Yoshihara}}, \bibinfo {author} {\bibfnamefont {T.}~\bibnamefont {Fuse}},
  \bibinfo {author} {\bibfnamefont {Z.}~\bibnamefont {Ao}}, \bibinfo {author}
  {\bibfnamefont {S.}~\bibnamefont {Ashhab}}, \bibinfo {author} {\bibfnamefont
  {K.}~\bibnamefont {Kakuyanagi}}, \bibinfo {author} {\bibfnamefont
  {S.}~\bibnamefont {Saito}}, \bibinfo {author} {\bibfnamefont
  {T.}~\bibnamefont {Aoki}}, \bibinfo {author} {\bibfnamefont {K.}~\bibnamefont
  {Koshino}},\ and\ \bibinfo {author} {\bibfnamefont {K.}~\bibnamefont
  {Semba}},\ }\bibfield  {title} {\bibinfo {title} {Inversion of qubit energy
  levels in qubit-oscillator circuits in the deep-strong-coupling regime},\
  }\href {https://doi.org/10.1103/PhysRevLett.120.183601} {\bibfield  {journal}
  {\bibinfo  {journal} {Phys. Rev. Lett.}\ }\textbf {\bibinfo {volume} {120}},\
  \bibinfo {pages} {183601} (\bibinfo {year} {2018})}\BibitemShut {NoStop}%
\bibitem [{\citenamefont {Langford}\ \emph {et~al.}(2017)\citenamefont
  {Langford}, \citenamefont {Sagastizabal}, \citenamefont {Kounalakis},
  \citenamefont {Dickel}, \citenamefont {Bruno}, \citenamefont {Luthi},
  \citenamefont {Thoen}, \citenamefont {Endo},\ and\ \citenamefont
  {DiCarlo}}]{Langford2017}%
  \BibitemOpen
  \bibfield  {author} {\bibinfo {author} {\bibfnamefont {N.~K.}\ \bibnamefont
  {Langford}}, \bibinfo {author} {\bibfnamefont {R.}~\bibnamefont
  {Sagastizabal}}, \bibinfo {author} {\bibfnamefont {M.}~\bibnamefont
  {Kounalakis}}, \bibinfo {author} {\bibfnamefont {C.}~\bibnamefont {Dickel}},
  \bibinfo {author} {\bibfnamefont {A.}~\bibnamefont {Bruno}}, \bibinfo
  {author} {\bibfnamefont {F.}~\bibnamefont {Luthi}}, \bibinfo {author}
  {\bibfnamefont {D.~J.}\ \bibnamefont {Thoen}}, \bibinfo {author}
  {\bibfnamefont {A.}~\bibnamefont {Endo}},\ and\ \bibinfo {author}
  {\bibfnamefont {L.}~\bibnamefont {DiCarlo}},\ }\bibfield  {title} {\bibinfo
  {title} {Experimentally simulating the dynamics of quantum light and matter
  at deep-strong coupling},\ }\href
  {https://doi.org/10.1038/s41467-017-01061-x} {\bibfield  {journal} {\bibinfo
  {journal} {Nat. Commun.}\ }\textbf {\bibinfo {volume} {8}},\ \bibinfo {pages}
  {1715} (\bibinfo {year} {2017})}\BibitemShut {NoStop}%
\bibitem [{\citenamefont {Braumüller}\ \emph {et~al.}(2017)\citenamefont
  {Braumüller}, \citenamefont {Marthaler}, \citenamefont {Schneider},
  \citenamefont {Stehli}, \citenamefont {Rotzinger}, \citenamefont {Weides},\
  and\ \citenamefont {Ustinov}}]{Braumuller2017}%
  \BibitemOpen
  \bibfield  {author} {\bibinfo {author} {\bibfnamefont {J.}~\bibnamefont
  {Braumüller}}, \bibinfo {author} {\bibfnamefont {M.}~\bibnamefont
  {Marthaler}}, \bibinfo {author} {\bibfnamefont {A.}~\bibnamefont
  {Schneider}}, \bibinfo {author} {\bibfnamefont {A.}~\bibnamefont {Stehli}},
  \bibinfo {author} {\bibfnamefont {H.}~\bibnamefont {Rotzinger}}, \bibinfo
  {author} {\bibfnamefont {M.}~\bibnamefont {Weides}},\ and\ \bibinfo {author}
  {\bibfnamefont {A.~V.}\ \bibnamefont {Ustinov}},\ }\bibfield  {title}
  {\bibinfo {title} {Analog quantum simulation of the {Rabi} model in the
  ultra-strong coupling regime},\ }\href
  {https://doi.org/10.1038/s41467-017-00894-w} {\bibfield  {journal} {\bibinfo
  {journal} {Nat. Commun.}\ }\textbf {\bibinfo {volume} {8}},\ \bibinfo {pages}
  {779} (\bibinfo {year} {2017})}\BibitemShut {NoStop}%
\bibitem [{\citenamefont {Hwang}\ \emph {et~al.}(2018)\citenamefont {Hwang},
  \citenamefont {Rabl},\ and\ \citenamefont {Plenio}}]{Hwang2018}%
  \BibitemOpen
  \bibfield  {author} {\bibinfo {author} {\bibfnamefont {M.-J.}\ \bibnamefont
  {Hwang}}, \bibinfo {author} {\bibfnamefont {P.}~\bibnamefont {Rabl}},\ and\
  \bibinfo {author} {\bibfnamefont {M.~B.}\ \bibnamefont {Plenio}},\ }\bibfield
   {title} {\bibinfo {title} {Dissipative phase transition in the open quantum
  {Rabi} model},\ }\href {https://doi.org/10.1103/PhysRevA.97.013825}
  {\bibfield  {journal} {\bibinfo  {journal} {Phys. Rev. A}\ }\textbf {\bibinfo
  {volume} {97}},\ \bibinfo {pages} {013825} (\bibinfo {year}
  {2018})}\BibitemShut {NoStop}%
\bibitem [{\citenamefont {Zhu}\ \emph {et~al.}(2020)\citenamefont {Zhu},
  \citenamefont {Ping}, \citenamefont {Yang},\ and\ \citenamefont
  {Agarwal}}]{Zhu2020}%
  \BibitemOpen
  \bibfield  {author} {\bibinfo {author} {\bibfnamefont {C.~J.}\ \bibnamefont
  {Zhu}}, \bibinfo {author} {\bibfnamefont {L.~L.}\ \bibnamefont {Ping}},
  \bibinfo {author} {\bibfnamefont {Y.~P.}\ \bibnamefont {Yang}},\ and\
  \bibinfo {author} {\bibfnamefont {G.~S.}\ \bibnamefont {Agarwal}},\
  }\bibfield  {title} {\bibinfo {title} {Squeezed light induced symmetry
  breaking superradiant phase transition},\ }\href
  {https://doi.org/10.1103/PhysRevLett.124.073602} {\bibfield  {journal}
  {\bibinfo  {journal} {Phys. Rev. Lett.}\ }\textbf {\bibinfo {volume} {124}},\
  \bibinfo {pages} {073602} (\bibinfo {year} {2020})}\BibitemShut {NoStop}%
\bibitem [{\citenamefont {Cai}\ \emph {et~al.}(2022)\citenamefont {Cai},
  \citenamefont {Liu}, \citenamefont {Jiang}, \citenamefont {Wu}, \citenamefont
  {Mei}, \citenamefont {Zhao}, \citenamefont {He}, \citenamefont {Zhang},
  \citenamefont {Zhou},\ and\ \citenamefont {Duan}}]{Cai2022}%
  \BibitemOpen
  \bibfield  {author} {\bibinfo {author} {\bibfnamefont {M.-L.}\ \bibnamefont
  {Cai}}, \bibinfo {author} {\bibfnamefont {Z.-D.}\ \bibnamefont {Liu}},
  \bibinfo {author} {\bibfnamefont {Y.}~\bibnamefont {Jiang}}, \bibinfo
  {author} {\bibfnamefont {Y.-K.}\ \bibnamefont {Wu}}, \bibinfo {author}
  {\bibfnamefont {Q.-X.}\ \bibnamefont {Mei}}, \bibinfo {author} {\bibfnamefont
  {W.-D.}\ \bibnamefont {Zhao}}, \bibinfo {author} {\bibfnamefont
  {L.}~\bibnamefont {He}}, \bibinfo {author} {\bibfnamefont {X.}~\bibnamefont
  {Zhang}}, \bibinfo {author} {\bibfnamefont {Z.-C.}\ \bibnamefont {Zhou}},\
  and\ \bibinfo {author} {\bibfnamefont {L.-M.}\ \bibnamefont {Duan}},\
  }\bibfield  {title} {\bibinfo {title} {Probing a dissipative phase transition
  with a {Trapped} ion through reservoir engineering},\ }\href
  {https://doi.org/10.1088/0256-307X/39/2/020502} {\bibfield  {journal}
  {\bibinfo  {journal} {Chinese Phys. Lett.}\ }\textbf {\bibinfo {volume}
  {39}},\ \bibinfo {pages} {020502} (\bibinfo {year} {2022})}\BibitemShut
  {NoStop}%
\bibitem [{\citenamefont {Song}\ \emph {et~al.}(2017)\citenamefont {Song},
  \citenamefont {Zheng}, \citenamefont {Zhang}, \citenamefont {Xu},
  \citenamefont {Zhang}, \citenamefont {Guo}, \citenamefont {Liu},
  \citenamefont {Xu}, \citenamefont {Deng}, \citenamefont {Huang},
  \citenamefont {Zheng}, \citenamefont {Zhu},\ and\ \citenamefont
  {Wang}}]{song_2017_continuousvariable}%
  \BibitemOpen
  \bibfield  {author} {\bibinfo {author} {\bibfnamefont {C.}~\bibnamefont
  {Song}}, \bibinfo {author} {\bibfnamefont {S.-B.}\ \bibnamefont {Zheng}},
  \bibinfo {author} {\bibfnamefont {P.}~\bibnamefont {Zhang}}, \bibinfo
  {author} {\bibfnamefont {K.}~\bibnamefont {Xu}}, \bibinfo {author}
  {\bibfnamefont {L.}~\bibnamefont {Zhang}}, \bibinfo {author} {\bibfnamefont
  {Q.}~\bibnamefont {Guo}}, \bibinfo {author} {\bibfnamefont {W.}~\bibnamefont
  {Liu}}, \bibinfo {author} {\bibfnamefont {D.}~\bibnamefont {Xu}}, \bibinfo
  {author} {\bibfnamefont {H.}~\bibnamefont {Deng}}, \bibinfo {author}
  {\bibfnamefont {K.}~\bibnamefont {Huang}}, \bibinfo {author} {\bibfnamefont
  {D.}~\bibnamefont {Zheng}}, \bibinfo {author} {\bibfnamefont
  {X.}~\bibnamefont {Zhu}},\ and\ \bibinfo {author} {\bibfnamefont
  {H.}~\bibnamefont {Wang}},\ }\bibfield  {title} {\bibinfo {title}
  {Continuous-variable geometric phase and its manipulation for quantum
  computation in a superconducting circuit},\ }\href
  {https://doi.org/10.1038/s41467-017-01156-5} {\bibfield  {journal} {\bibinfo
  {journal} {Nat. Commun.}\ }\textbf {\bibinfo {volume} {8}},\ \bibinfo {pages}
  {1061} (\bibinfo {year} {2017})}\BibitemShut {NoStop}%
\bibitem [{\citenamefont {Blais}\ \emph {et~al.}(2004)\citenamefont {Blais},
  \citenamefont {Huang}, \citenamefont {Wallraff}, \citenamefont {Girvin},\
  and\ \citenamefont {Schoelkopf}}]{Blais2004}%
  \BibitemOpen
  \bibfield  {author} {\bibinfo {author} {\bibfnamefont {A.}~\bibnamefont
  {Blais}}, \bibinfo {author} {\bibfnamefont {R.-S.}\ \bibnamefont {Huang}},
  \bibinfo {author} {\bibfnamefont {A.}~\bibnamefont {Wallraff}}, \bibinfo
  {author} {\bibfnamefont {S.~M.}\ \bibnamefont {Girvin}},\ and\ \bibinfo
  {author} {\bibfnamefont {R.~J.}\ \bibnamefont {Schoelkopf}},\ }\bibfield
  {title} {\bibinfo {title} {Cavity quantum electrodynamics for superconducting
  electrical circuits: An architecture for quantum computation},\ }\href
  {https://doi.org/10.1103/PhysRevA.69.062320} {\bibfield  {journal} {\bibinfo
  {journal} {Phys. Rev. A}\ }\textbf {\bibinfo {volume} {69}},\ \bibinfo
  {pages} {062320} (\bibinfo {year} {2004})}\BibitemShut {NoStop}%
\bibitem [{\citenamefont {Wallraff}\ \emph {et~al.}(2004)\citenamefont
  {Wallraff}, \citenamefont {Schuster}, \citenamefont {Blais}, \citenamefont
  {Frunzio}, \citenamefont {Huang}, \citenamefont {Majer}, \citenamefont
  {Kumar}, \citenamefont {Girvin},\ and\ \citenamefont
  {Schoelkopf}}]{Wallraff2004}%
  \BibitemOpen
  \bibfield  {author} {\bibinfo {author} {\bibfnamefont {A.}~\bibnamefont
  {Wallraff}}, \bibinfo {author} {\bibfnamefont {D.~I.}\ \bibnamefont
  {Schuster}}, \bibinfo {author} {\bibfnamefont {A.}~\bibnamefont {Blais}},
  \bibinfo {author} {\bibfnamefont {L.}~\bibnamefont {Frunzio}}, \bibinfo
  {author} {\bibfnamefont {R.-S.}\ \bibnamefont {Huang}}, \bibinfo {author}
  {\bibfnamefont {J.}~\bibnamefont {Majer}}, \bibinfo {author} {\bibfnamefont
  {S.}~\bibnamefont {Kumar}}, \bibinfo {author} {\bibfnamefont {S.~M.}\
  \bibnamefont {Girvin}},\ and\ \bibinfo {author} {\bibfnamefont {R.~J.}\
  \bibnamefont {Schoelkopf}},\ }\bibfield  {title} {\bibinfo {title} {Strong
  coupling of a single photon to a superconducting qubit using circuit quantum
  electrodynamics},\ }\href {https://doi.org/10.1038/nature02851} {\bibfield
  {journal} {\bibinfo  {journal} {Nature}\ }\textbf {\bibinfo {volume} {431}},\
  \bibinfo {pages} {162} (\bibinfo {year} {2004})}\BibitemShut {NoStop}%
\end{thebibliography}%


\begin{thebibliography}{2}%
\makeatletter
\providecommand \@ifxundefined [1]{%
 \@ifx{#1\undefined}
}%
\providecommand \@ifnum [1]{%
 \ifnum #1\expandafter \@firstoftwo
 \else \expandafter \@secondoftwo
 \fi
}%
\providecommand \@ifx [1]{%
 \ifx #1\expandafter \@firstoftwo
 \else \expandafter \@secondoftwo
 \fi
}%
\providecommand \natexlab [1]{#1}%
\providecommand \enquote  [1]{``#1''}%
\providecommand \bibnamefont  [1]{#1}%
\providecommand \bibfnamefont [1]{#1}%
\providecommand \citenamefont [1]{#1}%
\providecommand \href@noop [0]{\@secondoftwo}%
\providecommand \href [0]{\begingroup \@sanitize@url \@href}%
\providecommand \@href[1]{\@@startlink{#1}\@@href}%
\providecommand \@@href[1]{\endgroup#1\@@endlink}%
\providecommand \@sanitize@url [0]{\catcode `\\12\catcode `\$12\catcode
  `\&12\catcode `\#12\catcode `\^12\catcode `\_12\catcode `\%12\relax}%
\providecommand \@@startlink[1]{}%
\providecommand \@@endlink[0]{}%
\providecommand \url  [0]{\begingroup\@sanitize@url \@url }%
\providecommand \@url [1]{\endgroup\@href {#1}{\urlprefix }}%
\providecommand \urlprefix  [0]{URL }%
\providecommand \Eprint [0]{\href }%
\providecommand \doibase [0]{https://doi.org/}%
\providecommand \selectlanguage [0]{\@gobble}%
\providecommand \bibinfo  [0]{\@secondoftwo}%
\providecommand \bibfield  [0]{\@secondoftwo}%
\providecommand \translation [1]{[#1]}%
\providecommand \BibitemOpen [0]{}%
\providecommand \bibitemStop [0]{}%
\providecommand \bibitemNoStop [0]{.\EOS\space}%
\providecommand \EOS [0]{\spacefactor3000\relax}%
\providecommand \BibitemShut  [1]{\csname bibitem#1\endcsname}%
\let\auto@bib@innerbib\@empty
%</preamble>
\bibitem [{\citenamefont {Zheng}\ \emph {et~al.}(2023)\citenamefont {Zheng},
  \citenamefont {Ning}, \citenamefont {Chen}, \citenamefont {L\"u},
  \citenamefont {Shen}, \citenamefont {Xu}, \citenamefont {Zhang},
  \citenamefont {Xu}, \citenamefont {Li}, \citenamefont {Xia}, \citenamefont
  {Wu}, \citenamefont {Yang}, \citenamefont {Miranowicz}, \citenamefont
  {Lambert}, \citenamefont {Zheng}, \citenamefont {Fan}, \citenamefont {Nori},\
  and\ \citenamefont {Zheng}}]{Zheng2022}%
  \BibitemOpen
  \bibfield  {author} {\bibinfo {author} {\bibfnamefont {R.-H.}\ \bibnamefont
  {Zheng}}, \bibinfo {author} {\bibfnamefont {W.}~\bibnamefont {Ning}},
  \bibinfo {author} {\bibfnamefont {Y.-H.}\ \bibnamefont {Chen}}, \bibinfo
  {author} {\bibfnamefont {J.-H.}\ \bibnamefont {L\"u}}, \bibinfo {author}
  {\bibfnamefont {L.-T.}\ \bibnamefont {Shen}}, \bibinfo {author}
  {\bibfnamefont {K.}~\bibnamefont {Xu}}, \bibinfo {author} {\bibfnamefont
  {Y.-R.}\ \bibnamefont {Zhang}}, \bibinfo {author} {\bibfnamefont
  {D.}~\bibnamefont {Xu}}, \bibinfo {author} {\bibfnamefont {H.}~\bibnamefont
  {Li}}, \bibinfo {author} {\bibfnamefont {Y.}~\bibnamefont {Xia}}, \bibinfo
  {author} {\bibfnamefont {F.}~\bibnamefont {Wu}}, \bibinfo {author}
  {\bibfnamefont {Z.-B.}\ \bibnamefont {Yang}}, \bibinfo {author}
  {\bibfnamefont {A.}~\bibnamefont {Miranowicz}}, \bibinfo {author}
  {\bibfnamefont {N.}~\bibnamefont {Lambert}}, \bibinfo {author} {\bibfnamefont
  {D.}~\bibnamefont {Zheng}}, \bibinfo {author} {\bibfnamefont
  {H.}~\bibnamefont {Fan}}, \bibinfo {author} {\bibfnamefont {F.}~\bibnamefont
  {Nori}},\ and\ \bibinfo {author} {\bibfnamefont {S.-B.}\ \bibnamefont
  {Zheng}},\ }\bibfield  {title} {\bibinfo {title} {Observation of a
  superradiant phase transition with emergent cat states},\ }\href
  {https://doi.org/10.1103/PhysRevLett.131.113601} {\bibfield  {journal}
  {\bibinfo  {journal} {Phys. Rev. Lett.}\ }\textbf {\bibinfo {volume} {131}},\
  \bibinfo {pages} {113601} (\bibinfo {year} {2023})}\BibitemShut {NoStop}%
\bibitem [{\citenamefont {Hwang}\ \emph {et~al.}(2018)\citenamefont {Hwang},
  \citenamefont {Rabl},\ and\ \citenamefont {Plenio}}]{Hwang2018}%
  \BibitemOpen
  \bibfield  {author} {\bibinfo {author} {\bibfnamefont {M.-J.}\ \bibnamefont
  {Hwang}}, \bibinfo {author} {\bibfnamefont {P.}~\bibnamefont {Rabl}},\ and\
  \bibinfo {author} {\bibfnamefont {M.~B.}\ \bibnamefont {Plenio}},\ }\bibfield
   {title} {\bibinfo {title} {Dissipative phase transition in the open quantum
  {Rabi} model},\ }\href {https://doi.org/10.1103/PhysRevA.97.013825}
  {\bibfield  {journal} {\bibinfo  {journal} {Phys. Rev. A}\ }\textbf {\bibinfo
  {volume} {97}},\ \bibinfo {pages} {013825} (\bibinfo {year}
  {2018})}\BibitemShut {NoStop}%
\end{thebibliography}%

\end{document}

\title{Supplementary Material for ``Observation of photonic dynamics in dissipative quantum Rabi models"}
\author{Wen Ning\orcidlink{0000-0001-7919-8811}}
\thanks{These authors contribute equally to this work.}
\author{Ri-Hua Zheng\orcidlink{0000-0002-1944-1573}}
\thanks{These authors contribute equally to this work.}
\email{E-mail: ruazheng@gmail.com}
\author{Jia-Hao L\"{u}\orcidlink{0009-0007-2578-2415}}
\author{Ken Chen\orcidlink{0009-0007-1849-5258}}
\author{Xin Zhu}
\author{Fan Wu\orcidlink{0000-0002-1279-2258}}
\author{Zhen-Biao Yang\orcidlink{0000-0002-3964-4714}}
\email{E-mail: zbyang@fzu.edu.cn}
\author{Shi-Biao Zheng\orcidlink{0000-0002-9405-4709}}
\email{E-mail: t96034@fzu.edu.cn}
\affiliation{Fujian Key Laboratory of Quantum Information and Quantum\\
    Optics, College of Physics and Information Engineering, Fuzhou University, Fuzhou, Fujian 350108, China}

\maketitle

\tableofcontents
%\section{Derivation of the effective quantum Rabi Hamiltonian}

\section{Engineering the Rabi model with controlled unitary-dissipative competition}\label{appendixA}
The system under investigation composed of a resonator of frequency $\omega_{r}$ coupled to a qubit. The excitation energy of the qubit is periodically modulated as ($\hbar = 1$)
\begin{eqnarray}
    \omega =\omega _{0}+\varepsilon \cos (\nu_1 t),
\end{eqnarray}
where $\omega_0$ corresponds to the mean excitation energy of the qubit, and $\varepsilon $ and $\nu_1$ respectively denote the amplitude and frequency of
the applied parametric modulation. In addition to this longitudinal
modulations, the qubit is transversely driven with two external fields with
frequencies $\omega _{1}$ and $\omega _{2}$, respectively. The system
dynamics is described by Hamiltonian%
\begin{eqnarray}
    H=H_{0}+H_{I},
\end{eqnarray}
where%
\begin{eqnarray}
    H_{0}=(\omega_{r}- \delta )a^{\dagger }a+[\omega _{0}+\varepsilon
    \cos (\nu_1 t)]\left\vert e\right\rangle \left\langle e\right\vert ,
\end{eqnarray}
\begin{eqnarray}
    H_{I}= \delta a^{\dagger }a +[\lambda a^{\dagger }+\Omega
    _{1}e^{i(\omega _{1}t+\varphi _{1})}+\Omega _{2}e^{i(\omega _{2}t+\varphi_{2})}]\sigma _{-}+ {\rm H.c.},
\end{eqnarray}
$a^{\dagger }$ and $a$ denote the creation and annihilation operators for
the photonic field that is coupled to the qubit with a strength $\lambda $, $%
\Omega _{j}$ and $\varphi _{j}$ ($j=1,2$) denote the amplitude and phase of
the transverse drives applied to the qubit, whose ground and excited states
are denoted as $\left\vert g\right\rangle $ and $\left\vert e\right\rangle $%
, and $\sigma _{-}=\left\vert g\right\rangle \left\langle e\right\vert $.
Set $\omega _{1}=\omega _{0}$, $\varphi _{1}=0$, $\varphi _{2}=\pi /2$, $\nu_1+\omega _{0}=\omega _{r}-\delta $, and $\delta ,\lambda \ll \nu_1 $.
Then the resonator is coupled to the qubits at the first upper sideband with the detuning $\delta $. Performing the transformation $%
e^{i\int_{0}^{t}H_{0}dt}$, we obtain the system Hamiltonian in the
interaction picture%
\begin{eqnarray}
    H_{I}^{\prime }= \delta a^{\dagger }a+e^{-i\mu \sin (\nu_1 t)}[\lambda
    e^{i\nu_1 t}a^{\dagger }+\Omega _{1}+i\Omega _{2}e^{i\Delta\omega _{2}t}]\sigma
    _{-}+{\rm H.c.},
\end{eqnarray}
where $\mu =\varepsilon /\nu_1 $ and $\Delta \omega _{2}=\omega _{2}-\omega _{0}\ll \nu_1 $.\ Using the Jacobi-Anger expansion

\begin{eqnarray}
    \exp \left[ i\mu \sin \left( \nu_1 t\right) \right] =\stackrel{\infty }{\mathrel{\mathop{\sum }\limits_{m=-\infty }}
    }J_{m}(\mu )\exp \left( im\nu_1 t\right) ,
\end{eqnarray}
with $J_{m}(\mu )$ being the $m$th Bessel function of the first kind, we obtain
\begin{small}
    \begin{eqnarray}
        H_{I}^{\prime }= \delta a^{\dagger }a+\stackrel{\infty }{
            \mathrel{\mathop{\sum }\limits_{m=-\infty }}}J_{m}(\mu )e^{-im\nu t}[\lambda e^{i\nu_1 t}a^{\dagger }+\Omega _{1}+i\Omega
        _{2}e^{i\Delta \omega _{2}t}]\sigma _{-}+{\rm H.c.}.
    \end{eqnarray}
\end{small}
With the fast oscillating terms being discarded, the system Hamiltonian in the interaction picture reduces to
\begin{eqnarray}
    H_{I}^{\prime } &=&\delta a^{\dagger }a   +\frac{1}{2}\{K\sigma
    _{x}+ 2\eta (X\sigma _{x}-Y\sigma _{y})+\Omega[e^{i\Delta \omega _{2}t}(\sigma
    _{y}+i\sigma _{x})+{\rm H.c.}]\} \\\nonumber
    &=& \delta a^{\dagger }a
    +\frac{1}{2}\{K\sigma _{x}+2\eta (X\sigma_{x}-Y\sigma _{y})+2\Omega[\cos (\Delta \omega _{2}t)\sigma _{y}-\sin (\Delta\omega _{2}t)\sigma _{x}]\},
\end{eqnarray}
where $K=2\Omega _{1}J_{0}(\mu )$, $\Omega=\Omega _{2}J_{0}(\mu )$, $\eta=\lambda J_{1}(\mu )/2$, $X=a^{\dagger }+a$, and $Y=i(a^{\dagger }-a)$, $%
\sigma _{x}=(\sigma _{-}+\sigma _{-}^{\dagger })$, and $\sigma _{y}=i(\sigma
_{-}-\sigma _{-}^{\dagger })$. Under the transformation $\exp (iK\sigma
_{x}t/2)$, the system Hamiltonian becomes%
\begin{eqnarray}
    H_{I}^{^{\prime \prime }} &=& \delta a^{\dagger }a+\Omega\{\cos
    (\Delta \omega _{2}t)[\cos (Kt)\sigma _{y}-\sin (Kt)\sigma _{z}]-\sin
    (\Delta \omega _{2}t)\sigma _{x}\}
    +\eta \{X\sigma _{x}-Y[\cos (Kt)\sigma _{y}-\sin (Kt)\sigma
    _{z}]\}.
\end{eqnarray}
Under the conditions $\Delta \omega _{2}=$ $K\gg \eta ,\Omega$, the fast oscillating terms can be neglected \cite{Zheng2022}, and $H_{I}^{\prime \prime }$ reduces to the Rabi Hamiltonian%
\begin{eqnarray} \label{QRM1}
    H_{\rm Rabi}= \frac{\Omega}{2} \sigma_y+ \delta a^\dag a +\eta \sigma_x (a+ a^\dag).
\end{eqnarray}
Note $\sigma_{x(y)}$ are the Pauli operators of the qubit under the ground and exited states basis \{$|g\rangle$, $|e\rangle$\} and $\sigma_y$ can be treated as $\sigma_z$ here, after a simple representation transformation $S^\dag \sigma_y S$ with
\begin{eqnarray}
    S=\left(
    \begin{array}{cc}
        \frac{1}{\sqrt{2}} & \frac{i}{\sqrt{2}} \\
        \frac{i}{\sqrt{2}} & \frac{1}{\sqrt{2}}\\
    \end{array}
    \right).
\end{eqnarray}
With this implementation, the transverse drives have a two-fold decoupling
effect to suppress the unwanted longitudinal qubit-resonator coupling due to
the off-resonant coupling terms, which takes the form of $ka^{\dagger
}a\sigma _{z}$. Under the condition $ka^{\dagger }a\ll \Omega _{1}$, the
first driving effectively transforms such a longitudinal coupling
to $\frac{k^{2}}{2\Omega _{1}}(a^{\dagger }a)^{2}\sigma _{x}$, which is
further transformed by the second drive to $[\frac{k^{2}}{2\Omega _{1}}(a^{\dagger }a)^{2}]^{2}\sigma _{x}/\Omega$ when $\frac{k^{2}}{2\Omega _{1}}(a^{\dagger }a)^{2}\ll \Omega$.

\begin{figure}
    \centering
    \includegraphics[width=8.6cm]{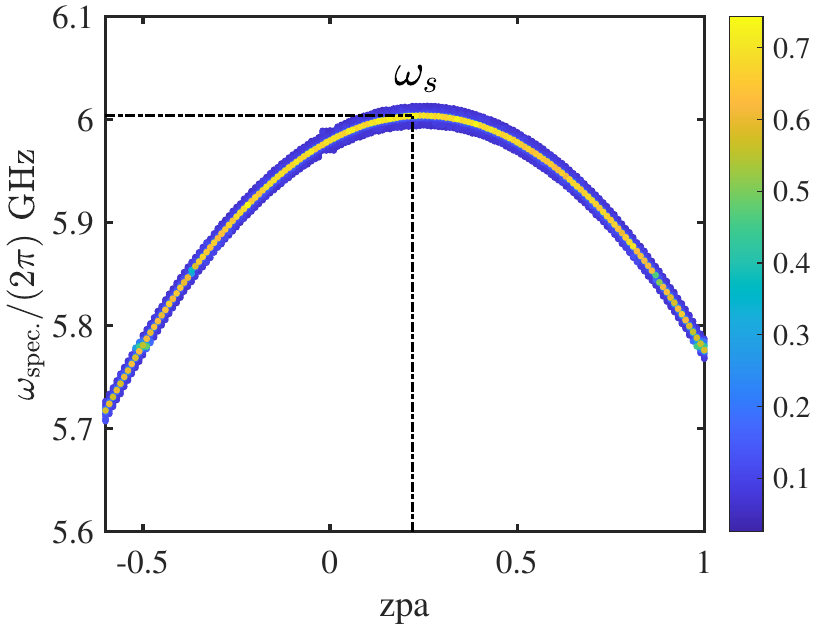}
    \caption{\textbf{Spectroscopy of the qubit frequency.}
        We apply square pulses with voltage amplitudes zpa ($x$-axis) to the Z line of the qubit.
        Meanwhile, several square-envelope pulses with certain amplitudes and specific frequencies $\omega_{\rm spec.}$ ($y$-axis) are applied to the XY line of the qubit to excite the qubit.
        Subsequently, we measure the populations of the qubit ($z$-axis).
        This spectroscopy reflects the relationship between the voltage amplitudes zpa and the qubit frequency.}
    \label{plt_spec_zpa}
\end{figure}

\begin{figure*}
    \centering
    \includegraphics[width=16cm]{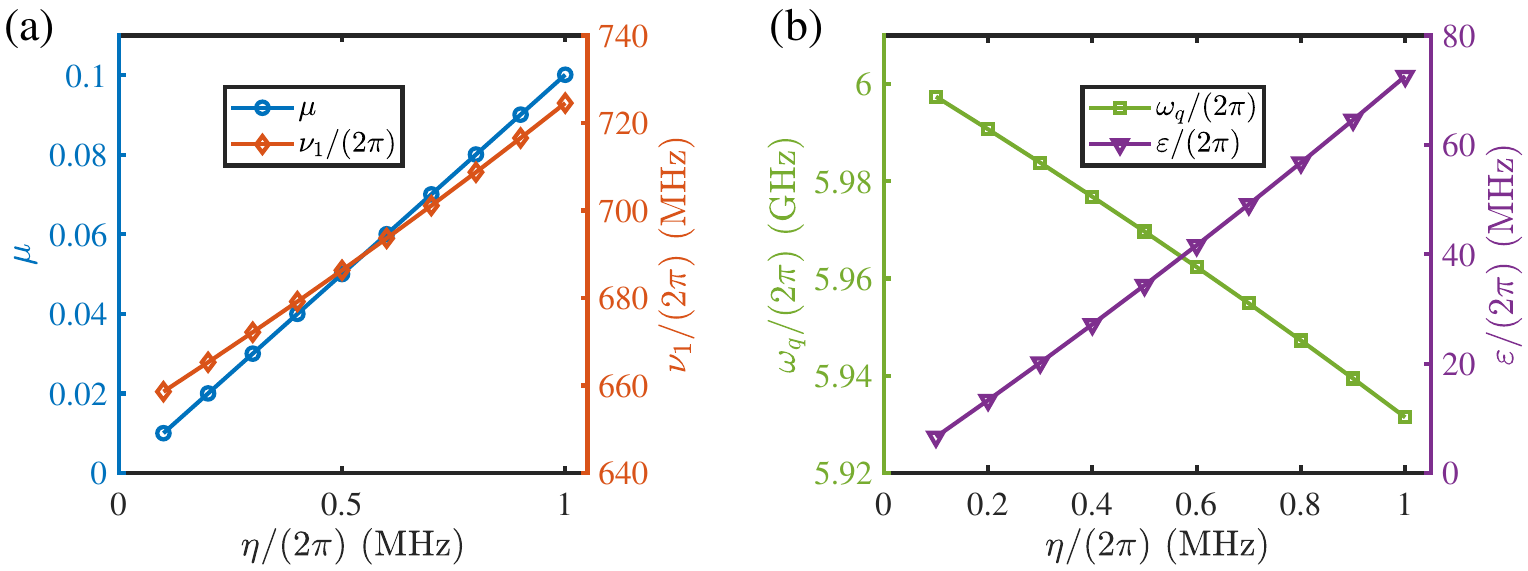}
    \caption{\textbf{Parameters versus effective coupling strength $\bm{ \eta/(2\pi)}$.}
        (a) Ratio of modulation amplitude to frequency $\mu$ and modulation frequency $\nu_1/(2\pi)$ versus effective coupling strength $\eta/(2\pi)$.
        (b) Qubit frequency, $\omega_q /(2\pi)$, and modulation amplitude $\varepsilon/(2\pi)$ versus effective coupling strength $\eta/(2\pi)$.
        Since $\mu$ is small, the results of the above parameters under Bessel function modulation tend to be linear.}
    \label{plt_paras}
\end{figure*}

The dissipative QRM dynamics can be described by the master equation
\begin{eqnarray}\label{master}
    \dot \rho=-i[H_{\rm Rabi},\rho]+\kappa a \rho a^\dag-\frac{\kappa}{2}(a^\dag a \rho +\rho a^\dag a),
\end{eqnarray}
where $\rho$ the density matrix, $a$ ($a^\dag$) the annihilation (creation) operator of the dissipative cavity mode with decay rate $\kappa$, and $H_{\rm Rabi}$ the QRM Hamiltonian.
The controlled unitary-dissipative competition of the QRM can be observed when the cavity decay rate $\kappa=5$ MHz is close to the coupling strength $\eta\sim2\pi \times 1 $ MHz, and further leads to a steady state with a stable photon number.
In this experiment, due to the large difference between cavity frequency and qubit frequency ($\sim 2\pi \times 700$ MHz), we have to perform the experiment near the sweet frequency [$\omega_s/(2\pi)= \times$ 6.004 GHz] of the qubit.
Specifically, since the relationship between the voltage zpa we apply to the z line and the qubit frequency $\omega_q$ is like an open downward quadratic function (see Fig. \ref{plt_spec_zpa}), we can modulate the periodic function near the zpa corresponding to the sweet frequency $\omega_s$ (i.e., the maximum value of this quadratic function).
This doubles the frequency of our input zpa and halves the amplitude.
In this mechanism, the qubit frequency $\omega_q$ needs to satisfy both relationships:
\begin{subequations}\label{relation1}
    \begin{align}
        \omega_q&=\omega_r-\nu_1, \\
        \omega_q&=\omega_s-\varepsilon,
    \end{align}
\end{subequations}
where $\omega_r/(2\pi)=6.656$ GHz is the frequency of the dissipative cavity.
The above two relationships in Eq. (\ref{relation1}) naturally form a constraint (note $\varepsilon=\mu\nu_1$)
\begin{eqnarray} \label{relation2}
    \nu_1=(\omega_r-\omega_s)/(1-\mu).
\end{eqnarray}
Based on the constraint in Eq. (\ref{relation2}), one can obtain a series of experimental parameters $\mu,\nu_1,\omega_q,\varepsilon$ with different effective coupling strengths $\eta/(2\pi)$ (see Fig. \ref{plt_paras}) by relation $\eta=\lambda J_1(\varepsilon/\nu_1)/2$.

\section{Simulation of the system dynamics} \label{AppendixB}
\begin{figure*}
    \centering
    \includegraphics[width=16cm]{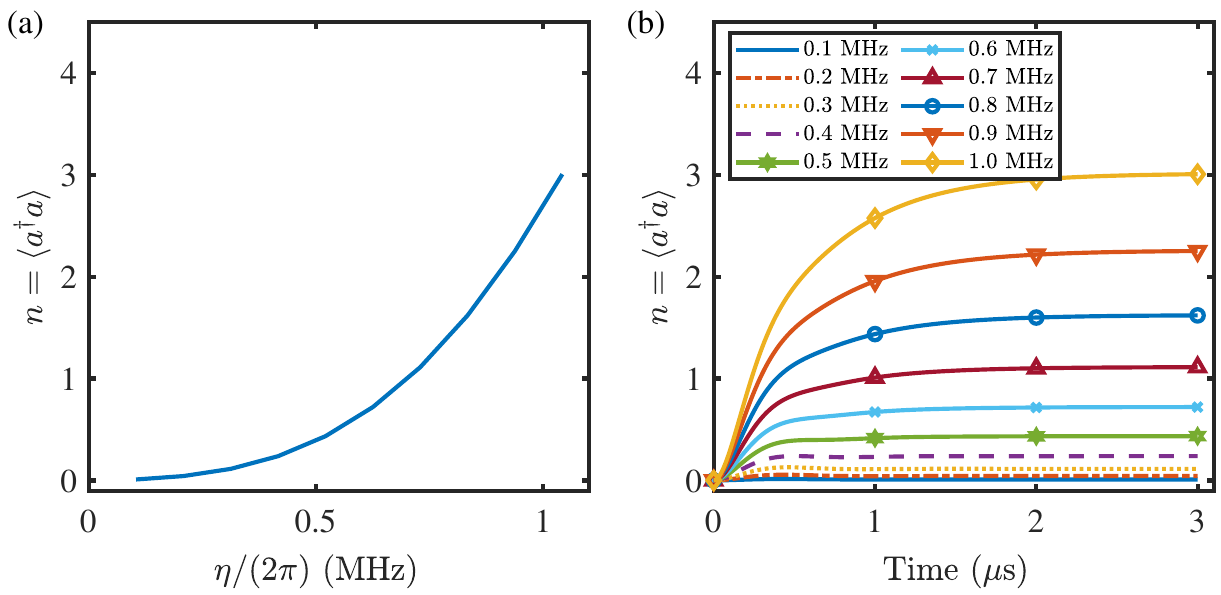}
    \caption{\textbf{Numerical simulation of the dissipative QRM dynamics. }
        (a) Photon number $n=\langle a^\dag a\rangle$ versus the QRM coupling $\eta/(2\pi)$.
        Each data point is simulated at $t=3$ $\rm \mu$s under the dissipative QRM dynamics with corresponding coupling $\eta/(2\pi)$.
        (b) The variation of the photon number $n=\langle a^\dag a\rangle$ during the dissipative QRM system's evolution towards the steady state with fixed coupling $\eta/(2\pi)=(0.1,0.2,0.3,0.4,0.5,0.6,0.7,0.8,0.9,1.0)$ MHz.	}
    \label{plt_sim}
\end{figure*}

\begin{figure*}
    \centering
    \includegraphics[width=16cm]{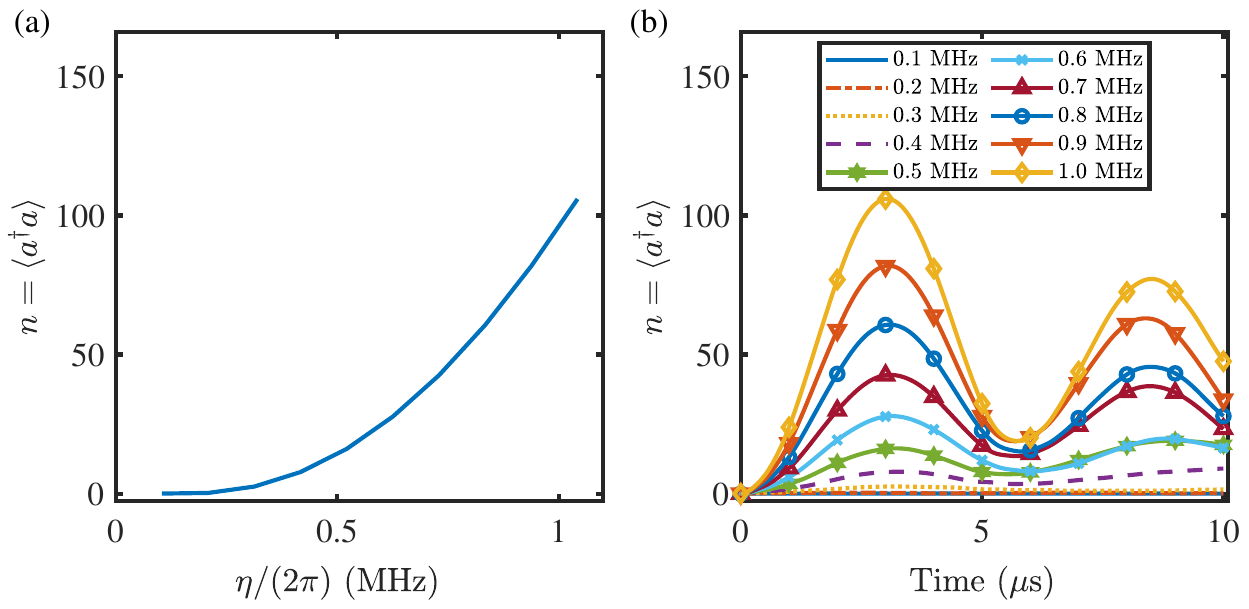}
    \caption{\textbf{Numerical simulation of the QRM dynamics without decoherence. }
        (a) Photon number $n=\langle a^\dag a\rangle$ versus the QRM coupling $\eta/(2\pi)$.
        Each data point is simulated at $t=3$ $\rm \mu$s under the QRM dynamics with corresponding coupling $\eta/(2\pi)$.
        (b) The variation of the photon number $n=\langle a^\dag a\rangle$ during the QRM system's evolution with fixed coupling $\eta/(2\pi)=(0.1,0.2,0.3,0.4,0.5,0.6,0.7,0.8,0.9,1.0)$ MHz.}
    \label{plt_sim_no_de}
\end{figure*}

The numerical results governed by the dissipative QRM dynamics in Eq. (\ref{master}) is shown in Fig. \ref{plt_sim}.
We chose parameters in Fig. \ref{plt_paras} to carry out the simulation with Eq. (\ref{master}) and $\delta/(2\pi)=$ 0.18 MHz.
Some of the above simulated results have been compared with the experimental data in the text, showing a relatively high consistency.
We also plot the results without decoherence in Fig. \ref{plt_sim_no_de}.
These results prove that the cavity dissipation plays an important part in the dynamics, making the system tend to a steady state, in stark contrast with the large number of photons production (see Fig. \ref{plt_sim_no_de}) by the unitary QRM without considering decoherence.
The results also show that the closer $\eta/(2\pi)$is to 1 MHz, the greater the rate of change of the steady-state photon number relative to $\eta$.
Because of the limitations of system parameters, the effective frequency ratio of the QRM is 5.6.
With the improvement of this ratio, it is possible to observe the dissipative superradiation phase transition, which was predicted to occur at the critical point $\xi_c=\sqrt{1+\kappa^2/(4\omega^2)}$ \cite{Hwang2018}, where the ratio approaches infinity.

\section{Observation of the photon number evolution} \label{signal}
\begin{figure*}
    \centering
    \includegraphics[width=16cm]{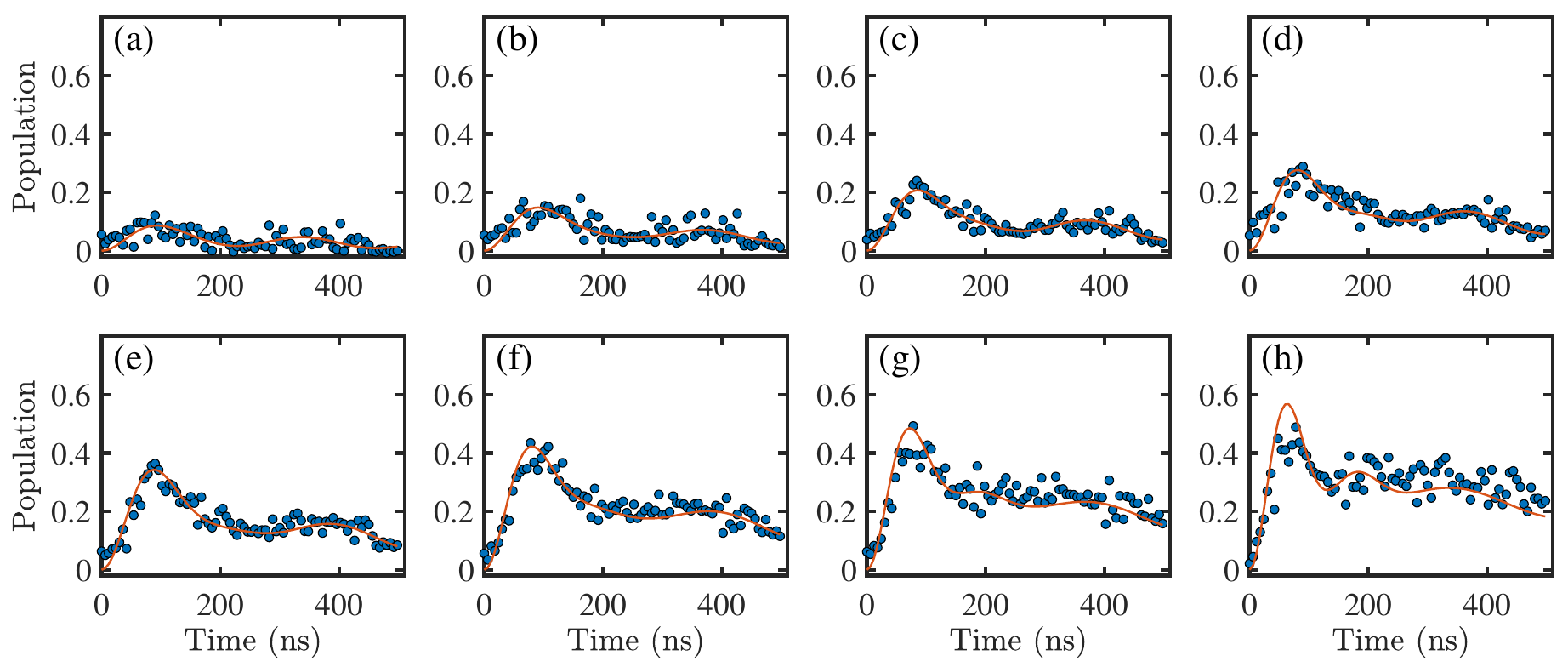}
    \caption{\textbf{Quantum Rabi oscillations for measuring the photon-number distribution.}
        (a-h) Rabi-oscillation signals after the 3-$\mu$s dissipative QRM dynamics when $\eta/(2\pi)=(0.3,0.4,0.5,0.6,0.7,0.8,0.9,1.0)$ MHz.
        These results are measured with the qubit, whose excitation is transferred to the bus resonator after the quantum Rabi dynamics, following which it is coupled to the dissipative resonator to extract the photon-number populations.}
    \label{plt_all_signal}
\end{figure*}

All the measured average photon numbers in the main text are deduced from the photon-number distribution.
In the experiment, after carrying out the dissipative QRM dynamics, the microwave drive and the frequency modulations are switched off.
Subsequently the excitation of the qubit is transferred to the bus resonator through a swapping gate at the frequency of the bus resonator.
Then the qubit is tuned on resonance with the dissipative resonator (frequency $6.656$ GHz) at the first sideband by applying a longitudinal modulation $\varepsilon \cos (\nu_1t)$.
Furthermore, the qubit undergoes photon-number-dependent Rabi oscillations.
The populations $P_{e}(\tau)$, of the excited state of the qubit for a given interaction time $\tau$, are measured by biasing the qubit back to its idle frequency, where its state is read out.
The recorded time-resolved quantum Rabi oscillation signals can be fitted by the Rabi-oscillation master equation
\begin{eqnarray}\label{Pnfit1}
    \dot \rho=-i[H_{\rm Ro},\rho]+\kappa a \rho a^\dag-\frac{\kappa}{2}(a^\dag a \rho +\rho a^\dag a),
\end{eqnarray}
with
\begin{eqnarray}\label{Pnfit2}
    H_{\rm Ro}=\eta |g\rangle \langle e|_q a^\dag+ {\rm H.c.},
\end{eqnarray}
yielding a fitted density matrix $\rho$ and further qubit population $P_{e}^{\rm f}(\tau)={\rm Tr}(\rho |e\rangle \langle e|)$.
One can find the most suitable density matrix $\rho$ by seeking the minimum value of $|P_{e}^{\rm f}(\tau)-P_{e}(\tau)|$.
In Fig. \ref{plt_all_signal}, we show the measured and fitted Rabi-oscillation signals, which indicate a good coincidence.

\section{Steady-state photon-number distributions}\label{AppendixD}
\begin{figure*}
    \centering
    \includegraphics[width=16cm]{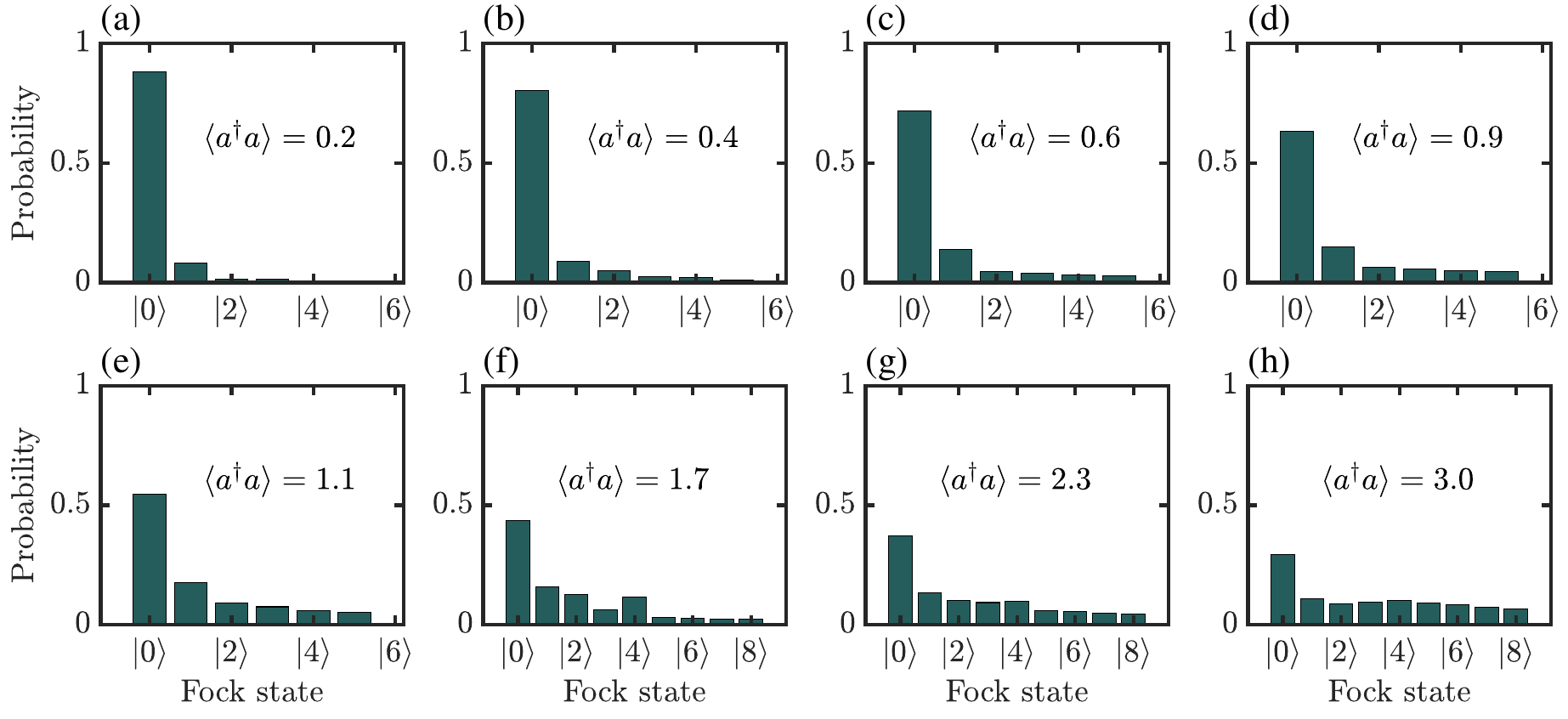}
    \caption{\textbf{Steady-state photon-number distributions.}
        (a-h) Photon-number distributions after the 3-$\mu$s dissipative QRM dynamics when $\eta/(2\pi)=(0.3,0.4,0.5,0.6,0.7,0.8,0.9,1.0)$ MHz.
        These results correspond to those in Fig. \ref{plt_all_signal}.}
    \label{plt_all_photon_dis}
\end{figure*}

After inferring the density matrix $\rho$, one can obtain the photon number distribution $P_n={\rm Tr}(\rho |n\rangle_c \langle n|)$ and the average photon number $n=\langle a^\dag a\rangle={\rm Tr}(\rho a^\dag a)$ in Fig. \ref{plt_all_photon_dis}.
The average photon numbers after the 3-$\mu$s dissipative QRM dynamics when $\eta/(2\pi)=(0.3,0.4,0.5,0.6,0.7,0.8,0.9,1.0)$ MHz are 0.2, 0.4, 0.6, 0.9, 1.1, 1.7, 2.3, and 3.0, respectively.
The results imply that the populations of relatively large photon numbers increase with the effective coupling strength.
For example, when $\eta/(2\pi)=0.8$ MHz the total population with three and more photons is 0.2684, which is increased to 0.4044 for $\eta/(2\pi)=0.9$ MHz.

\bibliography{ref}